\newif\ifShowKeys
\ShowKeysfalse

\newif\ifshowtikz
\showtikztrue
\showtikzfalse   % <---- comment/uncomment that line

\documentclass[11pt]{article}
	\pdfoutput=1
\usepackage[T1]{fontenc}

\usepackage{listings}
\lstnewenvironment{arkady}  % usage \begin{arkady} ... \end{arkady}
{\lstset{language=C,frame=trbl,basicstyle = \footnotesize \ttfamily , breaklines = true,showstringspaces=false}}{}

\usepackage{froufrou}
\usepackage{datetime}
\usepackage{comment}					% to comment large parts of text

\usepackage[setpagesize=false,pagebackref=false, 
linktocpage, bookmarksopen=true, colorlinks=true, 
linkcolor=blue,citecolor=blue,urlcolor=blue]{hyperref}

\usepackage[parsep]{collref}				% collect references in groups

\ifShowKeys \usepackage[notcite]{showkeys} \fi

\usepackage{amsmath, amssymb,amsthm}
\usepackage{stackrel}
\numberwithin{equation}{section}
\usepackage{bm,environ,mathrsfs,array,arydshln}
\usepackage{booktabs,float,slashed}
\usepackage{appendix}
\usepackage[mathcal]{euscript}
\usepackage{tensor} 						% Ratcliffe package to write tensors
\usepackage{mathabx}
\usepackage[vcentermath]{youngtab}
\usepackage{simpler-wick}

\usepackage{graphicx,epsfig,epic}
\usepackage{subcaption,wrapfig}
\usepackage{tikz}
\usepackage{tikz-feynman} 
\allowdisplaybreaks

\usepackage{framed}						% for shaded equations \begin{shaded} + \end{shaded} or \bs + \es
\definecolor{shadecolor}{rgb}{0.9996078, 0.984314, 0.960784}
\definecolor{framecolor}{rgb}{0,0,0}
\definecolor{TFTitleColor}{RGB}{1,1,1}
\definecolor{TFFrameColor}{RGB}{249	218	181}		
\definecolor{TFFrameColor}{RGB}{230 230 230 }

\newenvironment{frshaded}{%
    \MakeFramed {\FrameRestore}}%
    {\endMakeFramed}

\definecolor{myred}{RGB}{233, 33, 45}

\usepackage{mdframed}

\definecolor{lightpeach}{RGB}{255, 247, 235}

\newmdenv[
  backgroundcolor=gray!5,
  linecolor=gray!40,
  roundcorner=5pt,
  innerleftmargin=8pt,
  innerrightmargin=8pt,
  innertopmargin=6pt,
  innerbottommargin=6pt,
]{remarkbox}

\newenvironment{lab-remark}[1]{
  \begin{remarkbox}
  \textbf{\raisebox{0.2ex}{$\triangledown$} \textbf{ #1 } }\ignorespaces
}{
  \end{remarkbox}
}

\newcommand{\bs}{\begin{frshaded}}			% framed with background in shadecolor 
\newcommand{\es}{\end{frshaded}\noindent}

\def\ba#1\ea{\begin{align}#1\end{align}}		        %  clever way to bypass the known problem...
\newcommand{\be}{\begin{equation}}
\newcommand{\ee}{\end{equation}}
\newcommand{\bea}{\begin{equation} \begin{aligned}} 
\newcommand{\eea}{\end{aligned} \end{equation}}
\newcommand{\mc}{\mathcal }
\newcommand{\wh}{\widehat}
\newcommand{\wt}{\widetilde}

\newcommand{\mk}{\mathfrak}
\newcommand{\la}{\label}

\newcommand{\lp}{\notag \\ & }

\DeclareMathOperator{\tr}{\text{tr}}

\DeclareMathOperator{\LI}{Li}

\newcommand{\cf}{\textit{cf.} }
\newcommand{\ie}{\textit{i.e.} }

\newcommand{\K}{\mathsf{K}}
\newcommand{\FF}{{}_{2}F_{1}}
\newcommand{\PP}{\wt{\mc P}}

\newcommand{\sql}{\sqrt\l}
\renewcommand{\l}{\lambda}

\newcommand{\ket}[1]{|#1\rangle}
\newcommand{\bra}[1]{\langle #1|}
\newcommand{\braket}[2]{\langle #1|#2\rangle}
\newcommand{\mmm}[3]{\langle #1|#2|#3\rangle}

\newcommand{\sfa}{{\sf a}}
\newcommand{\sfb}{{\sf b}}

\begin{document}

\begin{comment}
\centerline{\Large\sc  Krylov Veneziano--Wosiek notes}
\vskip 0.2cm
\centerline{-- Notes --}
\vskip 0.2cm\centerline{\small\today\ -- \currenttime}
\vskip 0.5cm
 %\centerline{\sc M. Beccaria}
\bigskip
\begin{abstract}
\begin{center}
%\includegraphics[width=0.5\textwidth]{cover}
\end{center}
\end{abstract}
\end{comment}

%%%%%%%%%%%%%%%%%%%%%%%%%%%%%%%%%%%%%%%%%%%%%%%%%%%%%%%%%
%%%%% NOTICE the title page is commented out with \begin{comment}...\end{comment}   but is ready to be used

%\begin{comment}

\begin{titlepage}
%\begin{tabbing}
%\hspace*{11.5cm} \=  \kill % set the tabbings
%\>  Imperial-TP-AT-2024-?? \\
%\> %none
%\end{tabbing}

%\centerline{\small\today\ -- \currenttime}

\vspace*{15mm}
\begin{center}
{\Large\sc   Krylov Complexity in Supersymmetric \vskip 0.15cm  Large-$N$ Quantum Mechanics}%\vskip 9pt
%{\Large\sc      Notes}
\vspace*{10mm}

E. Alfinito$^{a}$ and M. Beccaria$^{a,b}$

\vspace*{4mm}
{\small
	
${}^a$ Universit\`a del Salento, Dipartimento di Matematica e Fisica \textit{Ennio De Giorgi},
\vskip 0.2cm
${}^{b}$ INFN - sezione di Lecce, Via Arnesano, I-73100 Lecce, Italy
\vskip 0.3cm
\vskip 0.2cm {\small E-mail: \texttt{matteo.beccaria@le.infn.it}}
}
\vspace*{0.8cm}
\end{center}

\begin{abstract}  
We study Krylov complexity in the large-$N$ planar limit of the 
supersymmetric matrix quantum mechanical  Veneziano--Wosiek model. In particular, we 
discuss the special features emerging at the critical transition at the 't~Hooft coupling $\l=1$.
Starting from selected states in the sectors with fermion number 0 and 1, related by supersymmetry, 
we analyze the time dependence of Krylov complexity by numerical methods. We find that  
for $\l\neq1$ the Krylov complexity $K(t)$ exhibits oscillatory behavior, while at the 
critical coupling $\l=1$ it grows quadratically in time, $K(t)\sim t^2$, with sector-dependent amplitudes.
To obtain analytical insight, 
we study in the bosonic sector a  solvable model with $\mk{sl}(2, \mathbb{R})$ symmetry which is 
a rank-1 modification of the Veneziano--Wosiek Hamiltonian, finding that it reproduces the previous features
of complexity. We also introduce supercharges and extend the solvable model to the fermionic sector where we also 
compute analytically the Krylov complexity. Higher degree-$M$ Krylov complexities, defined as expectation values of powers of Lanczos index, 
are also computed  and grow polynomially in time $\sim t^{2M}$ at the critical point both in the original and in the solvable model.
This behavior is closely analogous to the spreading of a localized squeezed state in a one-dimensional quantum harmonic oscillator of frequency $\omega$, 
with the free limit $\omega\to 0$ corresponding to the critical $\lambda\to 1$ limit.
\end{abstract}
\vskip 0.5cm
	{
		%Keywords: {\sc insert here keywords}
	}
\end{titlepage}

%\end{comment}
%%%%%%%%%%%%%%%%%%%%%%%%%%%%%%%%%%%%%%%%%%%%%%%%%%%%%%%%%

{\small
\makeatletter
\newcommand*{\toccontents}{\@starttoc{toc}}
\makeatother
\toccontents
}

%\tableofcontents

\vspace{1cm}

\setcounter{footnote}{0}

\section{Introduction and summary}

The growth of quantum complexity under time evolution provides a useful probe of 
scrambling and quantum chaos in many-body systems and has recently attracted renewed 
attention in the context of holography where it provides a microscopic candidate for gravitational complexity observables.
A  natural framework to characterize such 
dynamics is provided by the Krylov basis, generated by repeated action of the Hamiltonian 
on a reference operator or state through the Lanczos algorithm. \footnote{
In many systems with a spatial locality notion, quantum complexity may be examined by studying the spreading in time of a local operator $\mc O$ under unitary evolution. 
In the Heisenberg picture the evolved operator $\mc O(t)$ develops support over an increasingly large spatial region. Quantitative measures of 
this ``operator growth'' \cite{vonKeyserlingk:2017dyr,Khemani:2017nda} can be obtained from out-of-time-ordered correlators, which 
quantify the extent to which $\mc O(t)$ fails to commute with other simple local operators. Krylov complexity provides an alternative 
measure of operator growth which is conceptually simpler, as it depends only on the choice of inner product in operator space. In particular it may be 
defined without the need for an explicit spatial structure.
}
In this approach the dynamics is mapped to an effective tight-binding problem on a semi-infinite chain whose hopping amplitudes are the Lanczos coefficients
 \cite{Parker:2018yvk,Dymarsky:2019elm}, see  \cite{Nandy:2024evd,Rabinovici:2025otw} for recent reviews. 
%The 
%behavior of these coefficients has been argued to provide a diagnostic of chaotic dynamics, 
%with their asymptotic growth encoding universal features of operator spreading.
While 
originally formulated in the context of operator dynamics, the same construction can be 
applied to quantum states \cite{Balasubramanian:2022tpr,Hashimoto:2023swv}. \footnote{See also \cite{Caputa:2024vrn}
for quantification of complexity of states represented by density matrix operators.}
In this case, 
starting from an initial state $\ket{\psi_0}$, by successive application of the Hamiltonian
operator, one generates an 
orthonormal Krylov basis $\{\ket{K_n}\}$ and expands the time-evolved state as 
$\ket{\psi(t)} = \sum_n \phi_n(t)\,\ket{K_n}$. The spreading of the wavefunction along the Krylov 
chain provides a natural measure of dynamical complexity, quantified by the Krylov 
complexity
\be
\la{1.1}
K(t)=\sum_{n=0}^{\infty} n\,|\phi_n(t)|^2 .
\ee
General arguments relate the growth of $K(t)$, and thus operator spreading, to the asymptotic behavior of the Lanczos coefficients $\{\sfa_{n}, \sfb_{n}\}$. In particular, 
a linear growth $\sfb_n\sim n$ 
generically leads to exponential complexity growth characteristic of chaotic dynamics \cite{Parker:2018yvk,Rabinovici:2020ryf,Dymarsky:2019elm}; 
see also \cite{Muck:2022xfc}. 
\footnote{
\la{foot:3} For operator complexity the Lanczos states are generated by repeated action of the Liouvillian $[H,\cdot]$, 
which is anti-Hermitian with respect to the Hilbert–Schmidt inner product. As a result the diagonal Lanczos coefficients vanish identically. 
This simplification is absent for state complexity, where the diagonal coefficients are generically nonzero, 
making the classification of possible growth profiles less straightforward.
}

Applications of Krylov complexity to many-body systems and quantum field theories \cite{Dymarsky:2021bjq,Avdoshkin:2022xuw} have been widely explored. 
Examples include studies of the SYK model \cite{Stanford:2014jda,Jian:2020qpp,Jha:2024nbl,Ambrosini:2024sre,Xu:2024gfm,Chryssanthacopoulos:2025xyn,Anegawa:2024yia,
Bhattacharjee:2022ave,Chapman:2024pdw},
 random matrix theory \cite{Kar:2021nbm},  finite-size integrable systems \cite{Rabinovici:2021qqt}, transition from integrability to chaos \cite{Rabinovici:2022beu,Camargo:2023eev},
and confinement \cite{Jiang:2025wpj,Fatemiabhari:2026goj}.
More recently, Krylov complexity has been related to operator size growth and information scrambling in holographic settings, where it has been proposed as a probe of black hole dynamics and of the 
growth of quantum information behind horizons in AdS/CFT 
\cite{Susskind:2014rva,Rabinovici:2020ryf,Caputa:2018kdj,Jian:2020qpp,Caputa:2021sib,Kar:2021nbm,Lin:2022rbf,
Rabinovici:2023yex,Heller:2024ldz,Ambrosini:2024sre,Fu:2025kkh,Jeong:2026iac}. 
In particular, according to \cite{Susskind:2018tei,Caputa:2024sux,Fan:2024iop} 
the growth rate of Krylov complexity in holographic 
two-dimensional conformal field theories may be related to the proper radial momentum of a particle falling toward the horizon in the dual bulk geometry. 
\footnote{
If the geodesic motion avoids the interior singularity, the Krylov complexity may have an oscillatory behavior, as shown in \cite{zoakos2026} for the gravitational
description of complexity in the Coulomb branch of $\mc N=4$ SYM.
}
In this picture, operator spreading in the boundary theory is associated with motion along certain geodesics in the gravitational background, 
while the Krylov index represents an emergent (curved) dimension.
\footnote{
For further discussion of how microscopic measures of operator
growth may be related to near-horizon geometries see \cite{Magan:2018nmu,Lin:2019qwu,Kar:2021nbm,Iliesiu:2021ari}.
Krylov complexity has also been used to probe the properties of integrability breaking black hole–like states in spin chains and comparing them with 
integrable states \cite{Kristjansen:2025xqo}.}
Despite these 
developments, its behavior in simpler supersymmetric quantum mechanical systems remains 
comparatively unexplored, including models admitting a large-$N$ limit such as the 
Veneziano--Wosiek model, which provide a natural setting to investigate the interplay 
between supersymmetry, large-$N$ dynamics and quantum complexity.

The Veneziano--Wosiek model, originally introduced two decades ago,  is a supersymmetric matrix quantum mechanical model describing interacting bosonic and fermionic 
matrix degrees of freedom in the planar limit \cite{Veneziano:2005qs,Veneziano:2006dj}. The Hamiltonian is constructed from supersymmetry generators built out of 
bosonic and fermionic creation and annihilation operators carrying color indices. 
In the large-$N$ limit only planar contractions survive and the Hilbert space organizes into single-trace states, 
so that the dynamics reduces to an effective one-dimensional quantum mechanical problem with infinitely many degrees of freedom but highly constrained interactions. 
The model is supersymmetric and conserves fermion number,  and exhibits supersymmetric ground states with computable Witten index
\cite{Veneziano:2006bx,Veneziano:2006cx,Wosiek:2006tc,Onofri:2006hn,Trzetrzelewski:2007rr,Wosiek:2008bq,Korcyl:2007hw,Beccaria:2007re}.
At the critical value of the planar ’t~Hooft coupling, $\lambda=1$,  the energy eigenvalues collapse to zero and  the  
spectrum becomes increasingly dense. This signals a transition between two distinct regimes of the planar dynamics. 
\footnote{
The Veneziano–Wosiek model exhibits a weak–strong coupling duality relating the spectrum at coupling $\lambda$ to that at $1/\lambda$. 
In the planar basis the Hamiltonian reorganizes such that the interaction terms interchange the roles of splitting and joining processes 
under $\lambda \rightarrow 1/\lambda$. Consequently, the strong-coupling spectrum can be obtained from the weak-coupling 
one up to a simple rescaling of energies. This duality constrains the spectral structure across the transition near $\lambda=1$ 
and provides a nontrivial check on both analytical arguments and numerical computations.}
%For $\lambda \neq 1$ the planar spectrum is discrete and supersymmetry is realized with a finite number of 
%normalizable zero-energy states, jumping across $\l=1$. As the coupling approaches the critical value, the energy eigenvalues collapse to zero and  the  
%spectrum becomes increasingly dense. This signals a transition between two distinct regimes of the planar dynamics. 
%For $\lambda>1$ the low-energy sector is 
%dominated by configurations with large bosonic occupation number, and the Hilbert space effectively reorganizes around long single-trace states. 
%This behavior reflects the competition between the bosonic and fermionic contributions in the supersymmetric Hamiltonian 
%and has been interpreted as a phase transition in the large-$N$ quantum mechanical system.

Our motivation for revisiting this model is that it provides a particularly appealing setting to investigate Krylov complexity.
In fact, the dynamics of the model admits a remarkably simple representation in the planar single-trace basis, 
where the Hamiltonian acts locally  and takes a sparse, near-tridiagonal form,
reminiscent of a Krylov/Lanczos chain. \footnote{
In this sense, the Veneziano--Wosiek system offers a rare example in which the interplay between supersymmetry, 
large-$N$ dynamics and operator spreading can be explored in a controlled setting, see also \cite{Caputa:2021sib,Patramanis:2021lkx,Gamayun:2025hvu} 
for further exactly solvable cases. }
Schematically, if $\ket{n}$ denotes states ordered by bosonic occupation number, in a certain fermion number sector, 
the Hamiltonian acts as
\be
H\,\ket{n} = {\sf b}_{n+1}\ket{n+1} + {\sf a}_n \ket{n} + {\sf b}_n \ket{n-1},
\ee
so that the dynamics reduces to an effective tight-binding evolution along a one-dimensional chain.
The presence of a spectral transition at the critical coupling $\lambda=1$ suggests 
that different regimes of the theory may exhibit qualitatively different patterns of Krylov complexity growth. 
Studying the behavior of Krylov complexity in this model may therefore provide new insight into how 
supersymmetry, large-$N$ structure, and the emergence of a continuous spectrum
may influence the dynamical generation of complexity.

\subsection{Summary of results}

In this paper, we first analyze by numerical methods
the Veneziano--Wosiek model in the $F=0,1$ sectors, related by supersymmetry. Starting from special states 
in each sector, we evaluate the time dependence of the Krylov (spread) complexity, 
using a finite truncated Lanczos basis with dimension $\K$.
We pay special attention 
to the numerical convergence as the truncated Lanczos basis dimension $\K$ is increased. 

Besides the standard Krylov complexity defined in (\ref{1.1}), we will also consider its higher degree
generalization \cite{Muck:2022xfc} 
\be
K_{M}(t)=\sum_{n=0}^{\infty} n^{M}\,|\phi_n(t)|^2 .
\ee
Our analysis reveals two qualitatively distinct dynamical regimes. Away from the critical coupling the Krylov complexity remains bounded and exhibits oscillatory behavior, while at the critical point 
$\l=1$ it grows polynomially in time.

For $\l\neq 1$, the oscillations in Krylov complexity can be traced to the fact that the spectrum of the Veneziano--Wosiek model is asymptotically equally spaced. 
The time evolution  introduces asymptotically coherent phases %$e^{-iE_n t}$ 
and the spreading along the Krylov chain remains bounded.
In this case the Krylov complexity oscillates in time rather than growing indefinitely.

The behavior at the critical point $\l=1$ is qualitatively different. 
Here the  Krylov complexity is well described by a large $t$ quadratic growth $K^{F}_{M}(t)\sim C^{F}_{M} t^{2M}$, 
with an  amplitude $C^{F}_{M}$ that depends on the fermion-number sector. In other words, we observe
at criticality  a marginal non-exponential spreading of the wavefunction along the Krylov chain. 
This behavior appears to be closely related to the spectral rearrangement occurring 
at the critical coupling, where the planar dynamics becomes dominated by states with 
large bosonic occupation number. The critical point thus provides a natural setting 
in which the dynamics in Krylov space becomes effectively extended, leading to 
polynomial rather than oscillatory complexity growth.
\footnote{
Note that the quadratic growth observed at the critical point $\lambda=1$ does not appear to originate from a saturation of the Lanczos coefficients. 
As we will see, the leading asymptotic behaviour of the Lanczos coefficients is in fact independent of the planar coupling and 
the role of subleading terms is important.
}
%The evolution in Krylov space can be viewed as a tight-binding dynamics with hopping amplitudes $b_n$ and local potentials $a_n$. 
%If the typical Krylov index grows linearly in time, $n(t)\sim t$, the definition of Krylov complexity implies $K(t)\sim t^2$. 
%The behavior observed in the Veneziano--Wosiek model therefore suggests that at the critical coupling the 
%wavefunction spreads approximately linearly along the Krylov chain over an extended time interval. 
%This regime may be related to the spectral rearrangement at $\lambda=1$, 
%where states with large bosonic occupation number become increasingly important in the planar Hilbert space.

In order to obtain analytical insight, we introduce a companion model obtained from 
a rank-1 modification of the Veneziano--Wosiek Hamiltonian in the $F=0$ sector, 
first introduced  in \cite{DePietri:2006hi}. 
The model is solvable and belongs to a class of quantum systems, 
identified early on in \cite{Caputa:2021sib,Balasubramanian:2022tpr}, see also \cite{Rabinovici:2023yex,Caputa:2024sux}.
In these models, the Hamiltonian is built from the generators of
$\mk{sl}(2,\mathbb R)$ which plays the role of an emerging dynamical symmetry. \footnote{See \cite{Grabarits:2026hjz} for 
 exact results for arbitrary time-dependent generators.}
Building on standard methods to deal with these systems, one can obtain the exact analytical Krylov complexity
in the full Krylov space, \ie in $\K\to \infty$ limit. 
For $M=1$ one finds in the $F=0$ sector
\be
K^{F=0}_{1}(t; \l) = \frac{8\l}{(1-\l)^{2}}\sin^{2}\bigg(\frac{1}{2}(1-\l)\, t\bigg).
\ee
To extend the analysis to the fermionic sector $F=1$, we show that the bosonic $\mk{sl}(2, \mathbb R)$ model
admits explicit supercharges that allow to construct its 
supersymmetric partner Hamiltonian in the $F=1$ sector. We compute the exact analytical complexity 
also in this case and find the more involved result (again for $M=1$)
\be
K^{F=1}_{1}(t; \l) =  \frac{(1-\l)^{6}}{(1+\l)^{2}}\sum_{n=1}^{\infty}n\, (4\l)^{n}\,\frac{\sin^{2n}\mu\, \,(P_{1,n}(\l)^{2}\cos^{2}\mu+P_{2,n}(\l)^{2}\sin^{2}\mu)}{P_{3,n}(\l)\, (1+\l^{2}-2\l\cos(2\mu))^{n+3}},
\quad \mu\equiv\frac{1}{2}(1-\l)\,t .
\ee
Here $P_{1,n}(\l)$, $P_{2,n}(\l)$, and $P_{3,n}(\l)$ are polynomials in $\l$ and $\l^{-1}$ 
whose explicit form is given in the text. 
In both sectors, for $\l\neq 1$ the Krylov complexity oscillates in time: in the $F=0$ case it reduces 
to a simple $\sin^{2}$ oscillation. In the $F=1$ sector,  additional non-harmonic 
corrections appear.

The above exact expressions make it possible to determine the exact Krylov complexity at the critical 
point $\l=1$. In the $F=0$ and $F=1$ sectors, we find respectively
\be
\la{1.6}
K^{F=0}_{M}(t; 1) =  \sum_{k=1}^{M}(k+1)!\, S_{2}(M,k)\, t^{2k}, \qquad 
K^{F=1}_{M}(t; 1) =   \frac{1}{2}\sum_{k=1}^{M}(k+2)!\, S_{2}(M,k)\, t^{2k},
\ee
where $S_{2}(M,k)$ are Stirling numbers of the second kind. 
The Krylov complexity therefore grows as a polynomial in $t$ of degree $2M$ 
with positive integer coefficients. In particular, for $M=1$ one obtains
\be
K^{F=0}_{1}(t; 1) =  2\,t^{2}, \qquad
K^{F=1}_{1}(t; 1) =   3\,t^{2}.
\ee
The modified solvable model thus exhibits qualitative features very similar to those observed 
in the original Veneziano--Wosiek model. In particular, the change of dynamical regime 
at the critical point is reproduced, up to sector-dependent overall amplitudes. The reason for the agreement  
is that the rank-1 modification of the Veneziano--Wosiek model  does not change the asymptotic structure of the spectrum.

As a remark, the oscillatory behaviour of the Krylov complexity for $\lambda\neq 1$ and the simple structure in (\ref{1.6}) at $\l=1$ admit a simple 
physical interpretation in terms of a localized state in the quantum one-dimensional harmonic oscillator with varying angular frequency $\omega$.
Indeed, for a particle of unit mass and the initial state
\be
\braket{x}{\psi(0)} = \bigg(\frac{1}{2\pi\alpha}\bigg)^{1/4} e^{-\frac{x^{2}}{4\alpha}},
\ee
we prove in Section \ref{sec:HO} that
\be
K_{M}^{\rm HO}(t; \omega) =\frac{1}{\sqrt{\pi}}\sum_{k=1}^{M} S_{2}(M,k)\Gamma\left(k+\frac{1}{2}\right)\, \bigg(\frac{(1-4\alpha^{2}\omega^{2})^{2}}{16\alpha^{2}\omega^{2}}\sin^{2}\omega t\bigg)^{k}.
\ee
This expression oscillates for  generic $\omega$ and reduces in the free limit $\omega \to 0$ to the $2M$ degree polynomial
\be
K_{M}^{\rm HO}(t; 0) =\frac{1}{\sqrt{\pi}}\sum_{k=1}^{M} S_{2}(M,k)\Gamma\left(k+\frac{1}{2}\right)\, \bigg(\frac{t}{4\alpha}\bigg)^{2k},
\ee
which is qualitatively similar to (\ref{1.6}).

A number of interesting questions remain open. A first natural direction would be to extend the present analysis to higher fermion number sectors of the 
Veneziano--Wosiek model and to investigate whether similar analytic structures persist beyond the cases studied here. 
In particular, the critical point $\lambda=1$, where the discrete spectrum collapses and the dynamics becomes effectively 
governed by a continuous spectral density, deserves further investigation to clarify its relation to free quantum models 
at higher fermion number. 
One question is whether the polynomial growth of the complexity at the critical point 
may reflect a more general feature of quantum mechanical models with emergent continuum spectra. 

Another natural direction concerns 
the systematic analysis of the dependence of the complexity growth on the choice of the initial state and the extent to which the observed behavior 
is universal. Finally, it would be interesting to explore whether related solvable structures arise in other large-$N$ 
quantum mechanical or gauge models, where Krylov complexity may provide a useful probe of spectral properties, 
operator dynamics, and possible signatures of quantum chaos.

Finally, it would be interesting to examine whether
the supersymmetric structure of the 
modified solvable model admits a deeper algebraic interpretation. 
In the bosonic sector, the modified Hamiltonian is entirely characterized by the 
$\mk{sl}(2, \mathbb{R})$ symmetry algebra, whose representation theory governs 
both the spectrum and the Krylov complexity. The construction of the fermionic 
partner Hamiltonian suggests that 
the full supersymmetric model may be naturally embedded in an 
$\mk{osp}(1|2)$ superalgebra, the minimal supersymmetric extension of 
$\mk{sl}(2, \mathbb{R})$. If such an embedding exists, it would provide a unified 
algebraic framework for both sectors, potentially allowing the Krylov complexity 
in the fermionic sector to be derived directly from the representation theory of 
$\mk{osp}(1|2)$, in analogy with the bosonic case.

The paper is organized as follows. In Sec.~\ref{sec:krylov} we review the definition and basic properties of Krylov complexity. 
In Sec.~\ref{sec:VW-model} we introduce the Veneziano–Wosiek model and discuss the structure of its Hamiltonian in the large-$N$ limit. 
In Sec.~\ref{sec:VW-K} we analyze Krylov complexity in this model. 
In Sec.~\ref{sec:mod-model} we introduce a rank-1 modified Hamiltonian that preserves the tridiagonal structure and allows for analytic treatment. 
The corresponding Krylov complexity is computed analytically in Sec.~\ref{sec:mod-analytical} for a generic planar coupling $\l$.
In Sec.~\ref{sec:cont} we analyze the critical point $\lambda=1$, where the spectrum becomes continuous, and 
compute the complexity directly from the associated continuum eigenstates. 
Finally, in Sec.~\ref{sec:HO} we compute the Krylov complexity for a Gaussian state with generic width in the one-dimensional harmonic oscillator,
and show how the limit of vanishing angular frequency $\omega\to 0$  is similar to $\l\to 1$ in the Veneziano--Wosiek and modified models.
Additional technical details and further discussions are collected in the appendices.

\section{Krylov complexity}
\la{sec:krylov}

Krylov state complexity is defined in terms of the Lanczos basis. We start from 
a normalized state $\ket{s_{0}}$ and define
\be
\sfa_{0} = \mmm{s_{0}}{H}{s_{0}}, \qquad \sfb_{0}=0.
\ee
For $n=0, 1, \dots$, 
the Lanczos algorithm \footnote{For a discussion of numerical stability and improved algorithms see App. C in \cite{Rabinovici:2020ryf}.}
iterates the update 
\bea
\ket{z_{n+1}} &= (H-\sfa_{n})\ket{s_{n}}-\sfb_{n}\ket{s_{n-1}}, \\
\sfb_{n+1} &= \braket{z_{n+1}}{z_{n+1}}^{1/2}, \\
\ket{s_{n+1}} &= \sfb_{n+1}^{-1}\ket{z_{n+1}}, \\
\sfa_{n+1} &= \mmm{s_{n+1}}{H}{s_{n+1}}.
\eea
The Lanczos states $\ket{s_{n}}$  are orthonormal and obey
\be
\la{2.3}
H  \ket{s_{n}} = \sfa_{n} \ket{s_{n}}+\sfb_{n+1}\ket{s_{n+1}}+\sfb_{n}\ket{s_{n-1}}.
\ee
They correspond to a tridiagonal form of $H$ in the orthonormal Lanczos basis $\{|s_{n}\rangle\}$
that spans the Krylov space $\{H^{n}\ket{s_{0}}\}$
\be
H = \begin{pmatrix}
\sfa_{0} & \sfb_{1} \\
\sfb_{1} & \sfa_{1} & \sfb_{2} \\
           & \sfb_{2} & \sfa_{2} & \sfb_{3} & \\
&&\cdots        
\end{pmatrix}\, .
\ee
Once we write a generic state in the Lanczos basis
\be
\ket{\psi(t)} = \sum_{n=0}^{\infty}\psi_{n}(t)\, \ket{s_{n}},
\ee
the time evolution of the coefficients is 
\be
i\dot\psi_{n} = \sfb_{n}\psi_{n-1}+\sfa_{n}\psi_{n}+\sfb_{n+1}\psi_{n+1}, \qquad n=0, 1, 2, \dots,
\ee
with initial condition $\psi_{n}(0) = \delta_{n,0}$ and boundary condition $\psi_{-1}=0$. 
This may be truncated on a basis of $\K+1$ states by imposing $\psi_{\K+1}=0$.
The associated Krylov complexity is the matrix element
\be
\la{2.7}
K(t) = \sum_{n=1}^{\K}n|\psi_{n}(t)|^{2} = \mmm{s_{0}}{e^{-itH}\wh n e^{itH}}{s_{0}},
\ee
where $\wh n$ is the Krylov index diagonal operator $\wh n\, \ket{s_{n}} = n\, \ket{s_{n}}$. To make notation lighter, we will not denote
explicitly the basis dimension parameter $\K$. 
Although we will mainly be interested in the limit $\K\to\infty$, 
we will also briefly discuss finite-size effects at finite $\K$.

The Krylov complexity measures the average position of the evolving state along 
the Krylov chain. In this representation the Krylov dynamics is equivalent to a quantum walk 
on a one-dimensional chain with local potential $\sfa_{n}$ and hopping amplitudes $\sfb_{n}$.

\section{The Veneziano--Wosiek model}
\la{sec:VW-model}

The Veneziano--Wosiek model \cite{Veneziano:2005qs,Veneziano:2006dj} is built in terms of 
matrix creation and annihilation operators, where $T^{A}$ are the generators of $U(N)$ in the fundamental representation,
\bea
a_{ij} &=a^{A}T^{A}_{ij}, \qquad a^{\dag}_{ij}=a^{\dag A}T^{A}_{ij}, \qquad
f_{ij} =f^{A}T^{A}_{ij}, \qquad f^{\dag}_{ij}=f^{\dag A}T^{A}_{ij}, \\
A &=1, \dots, N^{2}, \qquad i,j=1, \dots N,
\eea
satisfying the algebra
\be
[a_{ij}, a^{\dag}_{kl}]=\{f_{ij}, f^{\dag}_{kl}\} = \delta_{il}\delta_{jk}.
\ee
The number operators are 
\be
B = \tr(a^{\dag}a), \qquad F=\tr(f^{\dag}f).
\ee
At large $N$ we will use  planar Fock states. In $F=0$ sector, these are 
\be
\ket{0; n} = \frac{1}{\mc N_{0,n}}\tr[(a^{\dag})^{n}]\, \ket{0},
\ee
where $\mc N_{0,n}$ is a normalization factor.
In $F=1$ sector we have instead
\be
\ket{1; n} =  \frac{1}{\mc N_{1,n}}\tr[(a^{\dag})^{n}f^{\dag}]\, \ket{0},
\ee
%In $F=2$ sector
%\be
%\ket{2; n_{1}, n_{2}} =  \frac{1}{\mc N_{2,n_{1},n_{2}}}\tr[(a^{\dag})^{n_{1}}f^{\dag}(a^{\dag})^{n_{2}}f^{\dag}]\, \ket{0}.
%\ee
%Due to cyclicity of the trace and the fermionic algebra one has
%\be
%\ket{2; n_{1}, n_{2}} = -\ket{2;n_{2},n_{1}}, \qquad \ket{2; n,n} = 0,
%\ee
and similar relations hold in higher sectors.
The Veneziano--Wosiek Hamiltonian is 
\bea
Q&=\tr[fa^{\dag}(1+ga^{\dag})], \qquad Q^{\dag}=\tr[f^{\dag}(1+ga)a], \\
H &= \{Q, Q^{\dag}\} = H_{B}+H_{F}, \\
H_{B} &= \tr[a^{\dag}a+g(a^{\dag 2}a+a^{\dag}a^{2})+g^{2}a^{\dag 2}a^{2}], \\
H_{F} &= \tr[f^{\dag}f+g(f^{\dag}f(a^{\dag}+a)+f^{\dag}(a^{\dag}+a)f)+g^{2}(f^{\dag}afa^{\dag}+f^{\dag}aa^{\dag}f
+f^{\dag}fa^{\dag}a+f^{\dag}a^{\dag}fa)],
\eea
and it conserves fermionic number.

The planar limit is defined by keeping $\l=g^{2}N$ fixed as $N\to\infty$. Let us consider the $F=0$ states $|0; n\rangle$.
States created by products of traces are subleading in the planar limit in the sense that they give rise to non-leading matrix elements. 
In the first $F=0,1$ sectors, the non-zero planar matrix elements are \cite{Veneziano:2005qs}
\bea
\la{3.7}
\mmm{0; n}{H}{0; n} &= (1+\l)n-\l\delta_{n,1}, \\
\mmm{0; n+1}{H}{0; n} &= \mmm{0; n}{H}{0; n+1} =  \sql\sqrt{n(n+1)}, \\
\eea
and %\footnote{Notice that this agrees with Eq.~(20) in \cite{Veneziano:2005qs}, but not with (15) and (16) there.}
\bea
\la{3.8}
\mmm{1; n}{H}{1; n} &=\l(n+2)+n+1,\qquad n \ge 0, \\
\mmm{1; n+1}{H}{1; n} &= \mmm{1; n}{H}{1; n+1} =  \sql(2+n).
\eea
These matrix elements can be reproduced by planar Hamiltonians written in terms of composite creation and annihilation operators 
\be
\la{3.9}
a^{\dag}_{n}\ket{0} = \ket{0; n},\ \ n\ge 0,  \qquad f^{\dag}_{n}\ket{0} = \ket{1; n-1},\ \ n\ge 1,
\ee
which satisfy the standard algebra
\be
[a_{n}, a^{\dag}_{m}] = \delta_{nm}, \qquad
\{f_{n}, f^{\dag}_{m}\} = \delta_{nm}.
\ee
We have indeed 
\ba
H^{F=0}&= a^{\dag}_{1}a_{1}+\sum_{n=2}^{\infty}n(1+\l)\, a^{\dag}_{n}a_{n}+(\sum_{n=1}^{\infty}\sql\sqrt{n(n+1)}\, a^{\dag}_{n}a_{n+1}+\text{h.c.}), \\
\la{3.12}
H^{F=1}&= \sum_{n=1}^{\infty}[n+(n+1)\l]f^{\dag}_{n}f_{n}+(\sum_{n=1}^{\infty}\sql\,(n+1)\,f^{\dag}_{n}f_{n+1}+\text{h.c.}).
\ea
% revised
The matrix elements in (\ref{3.7}), \ie in the $F=0$ sector, correspond to the Lanczos coefficients in (\ref{2.3})
\be
\sfa_{n} = (1+\l)\, n-\l\, \delta_{n,1}, \qquad \sfb_{n} = \sql\sqrt{n(n-1)}.
\ee
\paragraph{Remark} The coefficients $\sfa_{n}$ are non-vanishing as is typical for state (spread) complexity. We remark that in the study of operator Krylov
complexity one has $\sfa_{n}=0$ and the large time behavior of $K(t)$ is completely determined by the asymptotics of $\sfb_{n}$. In the presence
of non-zero diagonal coefficients this is no longer true and, in particular, $\sfb_{n}\sim n$ is not enough to imply an exponential growth of complexity.
As a simple example, one may consider the $\mk{sl}(2,\mathbb R)$ systems discussed in \cite{Caputa:2021sib} where the interplay between diagonal and off-diagonal
coefficients may produce exponentially growing or oscillating behaviour.

\subsection{Supersymmetry}

Let us introduce the supercharge $Q$ and its conjugate
\bea
\la{3.13}
Q &= a^{\dag}_{1}f_{1}+\sum_{n=1}^{\infty}\sqrt{n+1}\, a^{\dag}_{n+1}(f_{n+1}+\sql f_{n}), \\
Q^{\dag} &= f_{1}^{\dag}a_{1}+\sum_{n=1}^{\infty}\sqrt{n+1}\, (f^{\dag}_{n+1}+\sql f^{\dag}_{n})a_{n+1}.
\eea
The explicit action on states with $F=0, 1$ is 
\ba
Q \ket{0; n} &= 0, \\
Q \ket{1; n} & 
%\sum_{m=0}^{\infty}\sqrt{m+1}\, a^{\dag}_{m+1}f_{m+1}\ket{1; (n+1)-1}
%+\sql\sum_{m=1}^{\infty}\sqrt{m+1}\, a^{\dag}_{m+1} f_{m}\ket{1;(n+1)-1}\lp
%= \sum_{m=0}^{\infty}\sqrt{m+1}\, a^{\dag}_{m+1}f_{m+1}\ket{1; (n+1)-1}
%+\sql\sum_{m=0}^{\infty}\sqrt{m+2}\, a^{\dag}_{m+2} f_{m+1}\ket{1;(n+1)-1}\lp
= \sqrt{n+1}\ket{0; n+1}+\sql\sqrt{n+2}\ket{0; n+2},
\ea
and thus
\ba
Q^{\dag} \ket{1; n} &= 0, \\
Q^{\dag} \ket{0; n} &
%=\sum_{m=0}^{\infty}\sqrt{m+1}\, f^{\dag}_{m+1}a_{m+1}\ket{0; n}
%+\sql\sum_{m=1}^{\infty}\sqrt{m+1}\,  f^{\dag}_{m}a_{m+1}\ket{0; n}\lp
= \sqrt{n}\ket{1; n-1}+\sql \sqrt{n}\ket{1; n-2}.
\ea
The  supersymmetric algebra is 
\be
\{Q,Q\} = \{Q^{\dag},Q^{\dag}\}=0, \qquad \{Q, Q^{\dag}\}  = H,
\ee
where $H$ reduces to $H^{F}$ in the  $F=0,1$ sectors. As usual, as a consequence of the supersymmetry algebra, we have 
\be
[Q,H] = 0.
\ee

\subsection{Structure of the spectrum in $F=0,1$ sectors}

In the $F=0$ sector, we have the trivial vacuum $|0\rangle \equiv |0; 0\rangle$ with zero energy. 
There exists a second zero-energy state given by
\be
\la{3.20}
\ket{0}' = \sum_{n=1}^{\infty}(-1/\sql)^{n}\frac{1}{\sqrt n}\ket{0; n}.
\ee
This state is not normalizable for  $\l\le 1$ and is absent from the spectrum in that case. 

\begin{figure}[htbp]
    \centering

    \begin{subfigure}{0.4\linewidth}
        \centering
        \includegraphics[width=\linewidth]{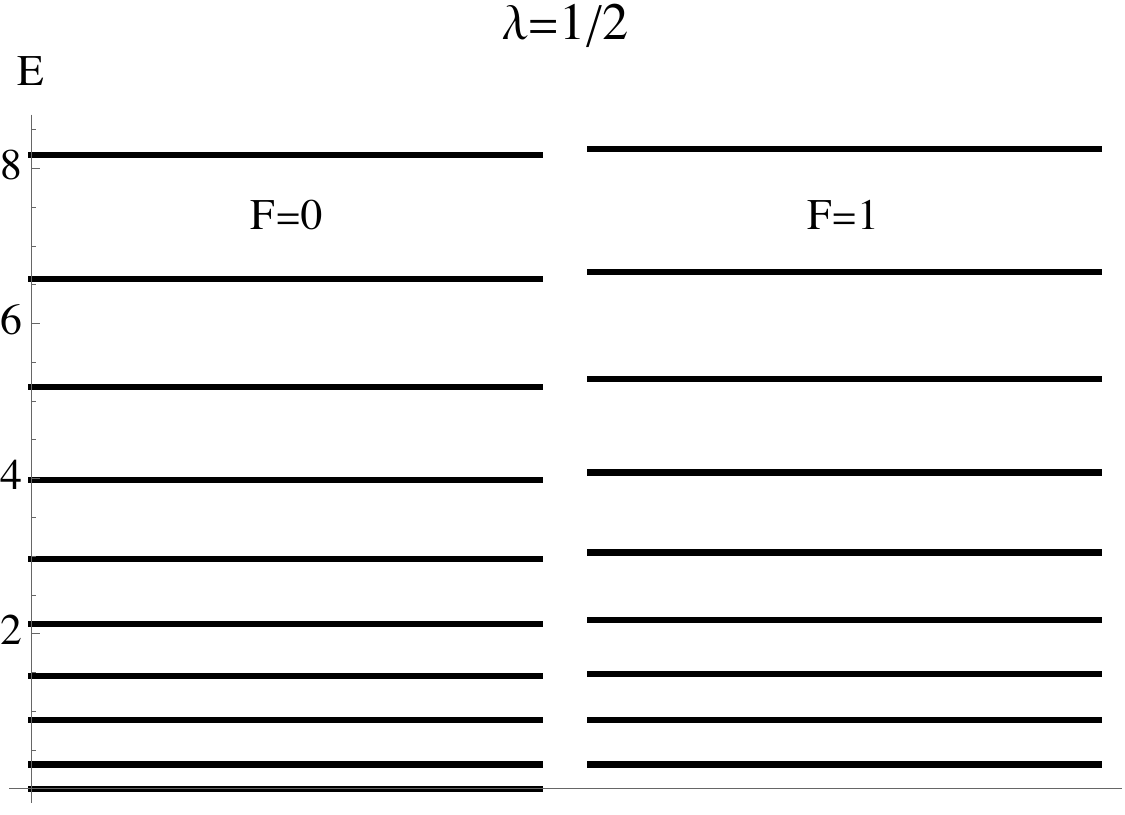}
        \caption{$N_{\rm bos}=20$}
%        \label{fig:sub1}
    \end{subfigure}
    \qquad
    \begin{subfigure}{0.4\linewidth}
        \centering
        \includegraphics[width=\linewidth]{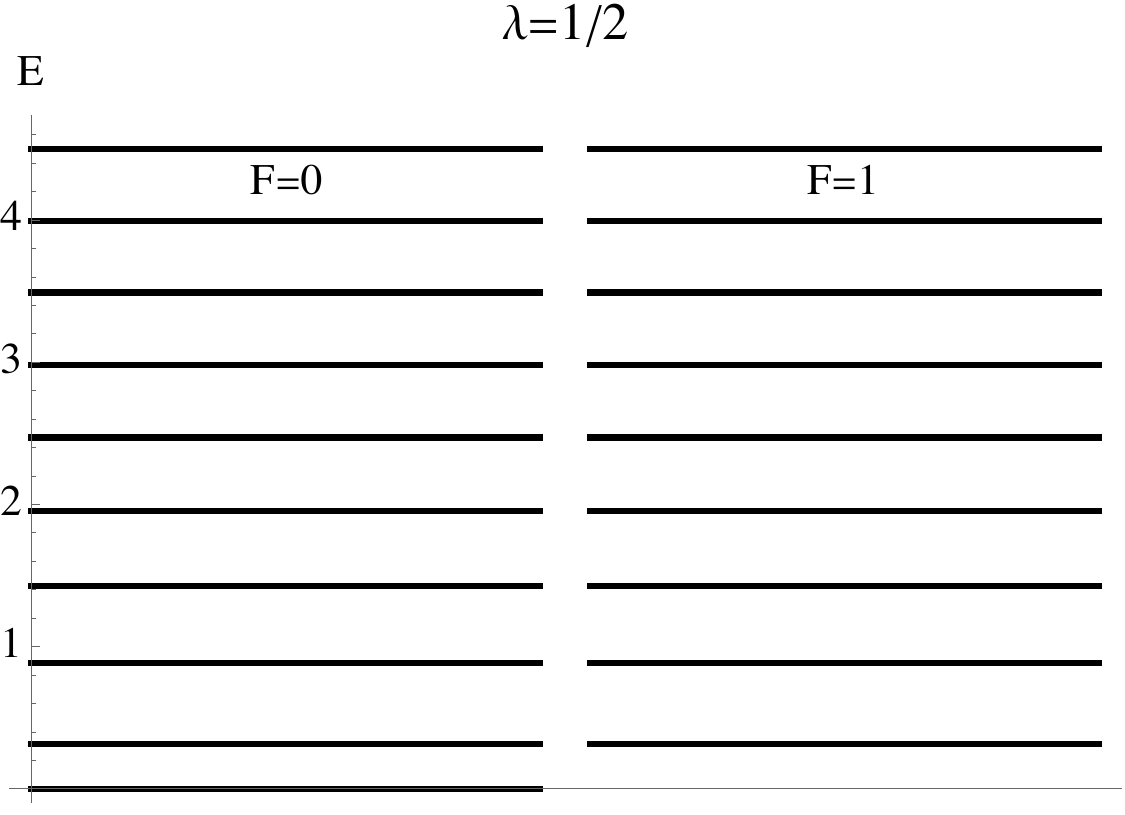}
        \caption{$N_{\rm bos}=100$}
%        \label{fig:sub2}
    \end{subfigure}

    \caption{Convergence of the spectrum of the planar Hamiltonians $H^{F=0}$ and $H^{F=1}$ at $\l=1/2$. Supersymmetric pairing of levels is 
    rather accurate in panel (b).}
    \label{fig:VW-spectrum}
\end{figure}

\noindent
The massive states are paired between the $F=0$ and $F=1$ sectors 
as can be verified numerically by truncating the Hilbert space at a sufficiently large
cut-off in the maximum number of bosonic excitations,
see Fig.~\ref{fig:VW-spectrum}.

At the critical point $\l=1$, all eigenvalues collapse to 0 as the cut-off is increased, approaching a continuous spectrum as the  cut-off becomes infinite.  
The exact spectrum in the $F=0$ sector was obtained in \cite{Veneziano:2005qs} for $\l\neq 1$ by noticing that 
\be
H^{F=0}=\sum_{n=1}^{\infty}B_{n}^{\dag}B_{n}, \qquad B_{n}=\sqrt n \, a_{n}+\sql\,\sqrt{n+1}\,a_{n+1},
\ee
where $a_{n}$ and $a^{\dag}_{n}$ annihilate and create states $\ket{n}\equiv \ket{0; n}$ with 
\be
[a_{n},a^{\dag}_{m}] = \delta_{n,m}.
\ee
The states \footnote{
The Krylov states introduced previously $\{\ket{s_{0}}, \ket{s_{1}}, \dots\} = \{\ket{1}, \ket{2}, \dots\}$
are related to this basis by
 the inverse relation 
$\ket{n} = \ket{s_{n-1}} =\frac{1}{\sqrt n} \sum_{p=0}^{\infty}(-\sql)^{p}\ket{B_{n+p}}$.
}
\be
\ket{B_{n}} = B^{\dag}_{n}\ket{0} = \sqrt n\, \ket{n}+\sql\, \sqrt{n+1}\, \ket{n+1},
\ee
are not orthonormal, but still $H^{F=0}$ is tridiagonal and its spectral problem can be solved.
%\ba
%\bar H \ket{B_{0}} &= -\sql \ket{B_{0}}+\ket{B_{1}}, \\
%\bar H \ket{B_{1}} &= \bigg(\sql+\frac{1}{\sql}\bigg) \ket{B_{1}}+2\ket{B_{2}}, \\
%\bar H \ket{B_{n}} &= n \ket{B_{n-1}}+n\bigg(\sql+\frac{1}{\sql}\bigg) \ket{B_{n}}+(n+1) \ket{B_{n+1}}, \qquad n\ge 2.
%\ea
A generic eigenvector can be written in the $\ket{B_{n}}$ basis
\be
\ket{\psi_{\alpha}} = \sum_{n=0}^{\infty}\psi_{\alpha, n}\ket{B_{n}}, 
\ee
with exact generating function
%\be
%f_{\alpha}(x) = \sum_{n=0}^{\infty}\psi_{\alpha, n}x^{n},
%\ee
%is found to be 
\be
f_{\alpha}(x) \equiv  \sum_{n=0}^{\infty}\psi_{\alpha, n}x^{n} = \begin{cases}
\frac{1}{\xi_{\alpha}}\frac{1}{x+1/\sql}\,\FF(1, \xi_{\alpha}, 1+\xi_{\alpha}; \frac{x+\sql}{x+1/\sql}), \qquad\qquad \l<1, \\
\frac{1}{1-\xi_{\alpha}}\frac{1}{x+\sql}\,\FF(1, 1-\xi_{\alpha}, 2-\xi_{\alpha}; \frac{x+1/\sql}{x+\sql}), \qquad \l>1.
\end{cases}
\qquad
\xi_{\alpha} =  \frac{E_{\alpha}}{\l-1},
\ee
The eigenvalues are determined by the quantization condition
\be
\la{3.26}
f_{\alpha}(0) = 0, \qquad
\begin{cases}
F(1, \xi_{\alpha}, 1+\xi_{\alpha}; \l) \quad\ \  \, = B(\l,\frac{E_{\alpha}}{\l-1}, 0)\ \  \ \  \ \, =0, \qquad \l<1, \\
F(1, 1-\xi_{\alpha}, 2-\xi_{\alpha}; \frac{1}{\l}) = B(\frac{1}{\l}, 1-\frac{E_{\alpha}}{\l-1},0)=0, \qquad \l>1,
\end{cases}
\ee
where $B(z, a,b)$ is the incomplete beta function.  The $\l\to 1$ limit can be extracted from this condition. For instance for $\l\to 1^{-}$ we find \footnote{We denote by $\psi(z)$ the
digamma function with $\psi(n+1)+\gamma_{\rm E} = \sum_{p=1}^{n}\frac{1}{p}$ for $n=0,1,2,\dots$.
}
\be
\la{3.27}
E_{n} = (1-\l)\bigg[n-\frac{1}{\log(1-\l)}+(\psi(n+1)+\gamma_{\rm E})\frac{1}{\log^{2}(1-\l)}+\cdots\bigg], \ \ n=0, 1, 2, \dots,
\ee
showing once again that as $\l\to1$ the eigenvalues collapse to zero.
Although the quantization condition (\ref{3.26}) can be easily solved numerically for any $\l$, 
the bosonic spectrum is non-trivial \footnote{The fermionic spectrum
in the $F=1$ sector is given by supersymmetric pairing as discussed previously, up to zero energy states.}
making it difficult to obtain a closed analytic expression for the Krylov complexity.

\section{Krylov complexity in the Veneziano--Wosiek model}
\la{sec:VW-K}

In the Veneziano--Wosiek model the fermion number is conserved and we introduce the notation 
\be
K^{F=0,1}(t; \l), \qquad K^{F=0,1}(t)\equiv K^{F=0,1}(t; 1).
\ee
The Krylov complexity depends on the initial state. 
A key observation is that the tridiagonal structure of $H^{F}$ implies that the special choice  
\be
\la{4.2}
F=0: \qquad \ket{s_{0}} = \ket{0,1}, \qquad
F=1: \qquad \ket{s_{0}} = \ket{1,0},
\ee
is such that the Krylov chain is simply
\be
F=0: \qquad \ket{s_{n}} = \ket{0,n+1}, \qquad
F=1: \qquad \ket{s_{n}} = \ket{1,n}.
\ee
If we denote by $\wh N$ the (bosonic) number operator such that $\wh N\, \ket{F, n} = n\ket{F,n}$, the relation between $\wh N$ and the Krylov index operator $\wh n$ 
is 
\be
F=0: \qquad \wh n = \wh N-1, \qquad
F=1: \qquad \wh n = \wh N.
\ee
In the following we will mainly adopt the choice (\ref{4.2}), which leads to
\bea
K^{F=0}(t; \l) &=\mmm{0; 1}{e^{-itH}(\wh N-1) e^{itH}}{0; 1}, \qquad
K^{F=1}(t; \l) =\mmm{1; 0}{e^{-itH}\wh N e^{itH}}{1; 0},
\eea
We also study the generalized higher order Krylov complexities ($K_{1}\equiv K$) defined by
\be
K_{M}(t) = \sum_{n=1}^{\K}n^{M}|\psi_{n}(t)|^{2}.
\ee
In our case, we have explicitly
\bea
K_{M}^{F=0}(t; \l) &=\mmm{0; 1}{e^{-itH}(\wh N-1)^{M} e^{itH}}{0; 1}, \qquad
K_{M}^{F=1}(t; \l) =\mmm{1; 0}{e^{-itH}\wh N^{M} e^{itH}}{1; 0}.
\eea

\subsection{Numerical analysis with truncated Lanczos basis}

\begin{figure}[htbp]
    \centering

    \begin{subfigure}{0.4\linewidth}
        \centering
        \includegraphics[width=\linewidth]{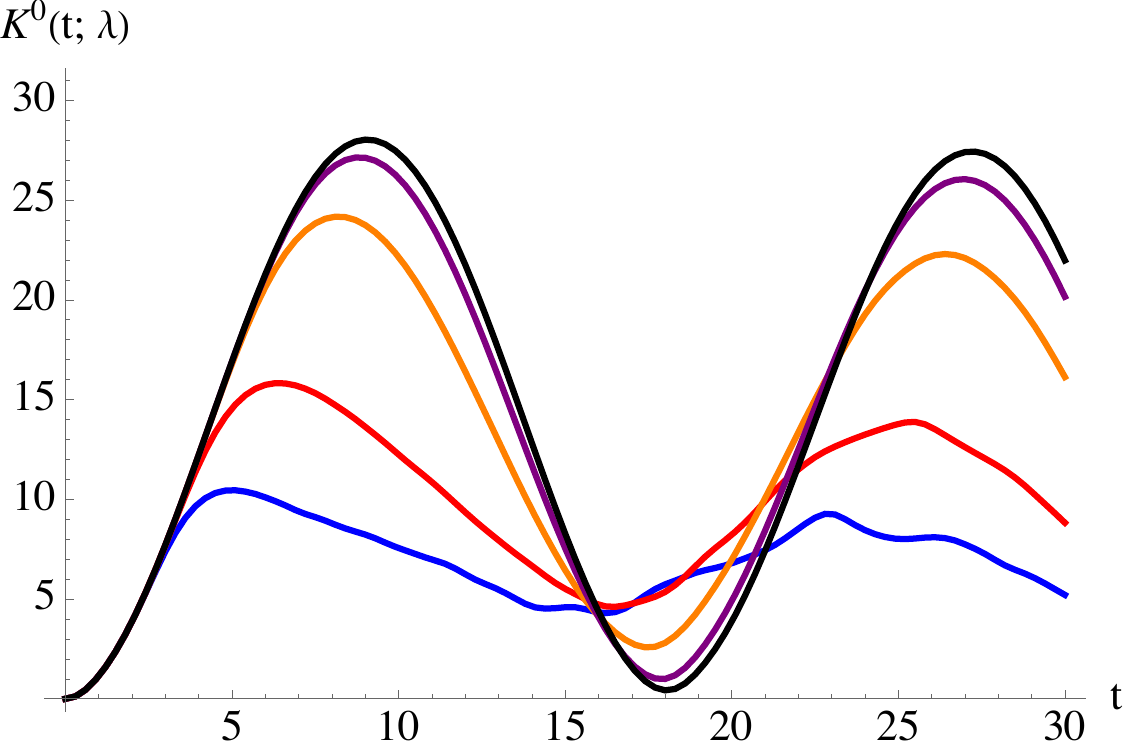}
        \caption{$\l=2/3$}
%        \label{fig:sub1}
    \end{subfigure}
    \qquad
    \begin{subfigure}{0.4\linewidth}
        \centering
        \includegraphics[width=\linewidth]{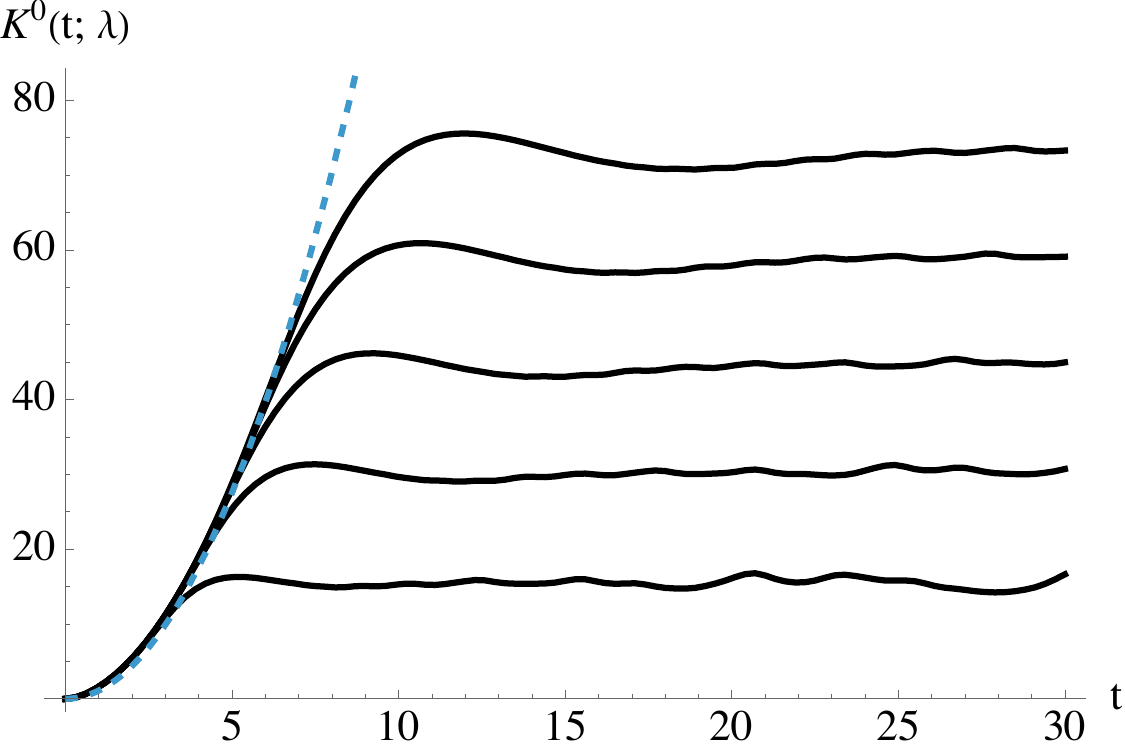}
        \caption{$\l=1$}
%        \label{fig:sub2}
    \end{subfigure}

    \caption{Numerical analysis of bosonic complexity $K^{0}(t; \l)$.  
    The left panel is for $\l=2/3$, a generic value smaller than 1. Blue, red, orange, purple, and black lines correspond to $\K=30,50,100,150,250$.
    Further increasing $\K$ does not significantly change the complexity in this temporal window. 
    The final complexity oscillates. Similar behaviour is observed for $\l>1$. In the right panel, we show results for $\l=1$. Curves from bottom to top
    correspond to $\K=50,100,150,200,250$. At the critical coupling, we see that 
    the complexity saturates at large times for any fixed $\K$, while the envelope of the curves grows as $\K$ increases and 
    grows approximately as $t^{2}$ (the dashed blue line  is $1.1\, t^{2}$).}
    \label{fig:VW-K0}
\end{figure}

We evaluate numerically the Krylov complexity $K^{F}(t;\l)$ at fixed $\l$ and increasing the Krylov space dimension $\K$. In Fig.~\ref{fig:VW-K0}
we show the result for $F=0$ and $\l=2/3$ (left) or $\l=1$ (right).

\begin{figure}[htbp]
    \centering

    \begin{subfigure}{0.4\linewidth}
        \centering
        \includegraphics[width=\linewidth]{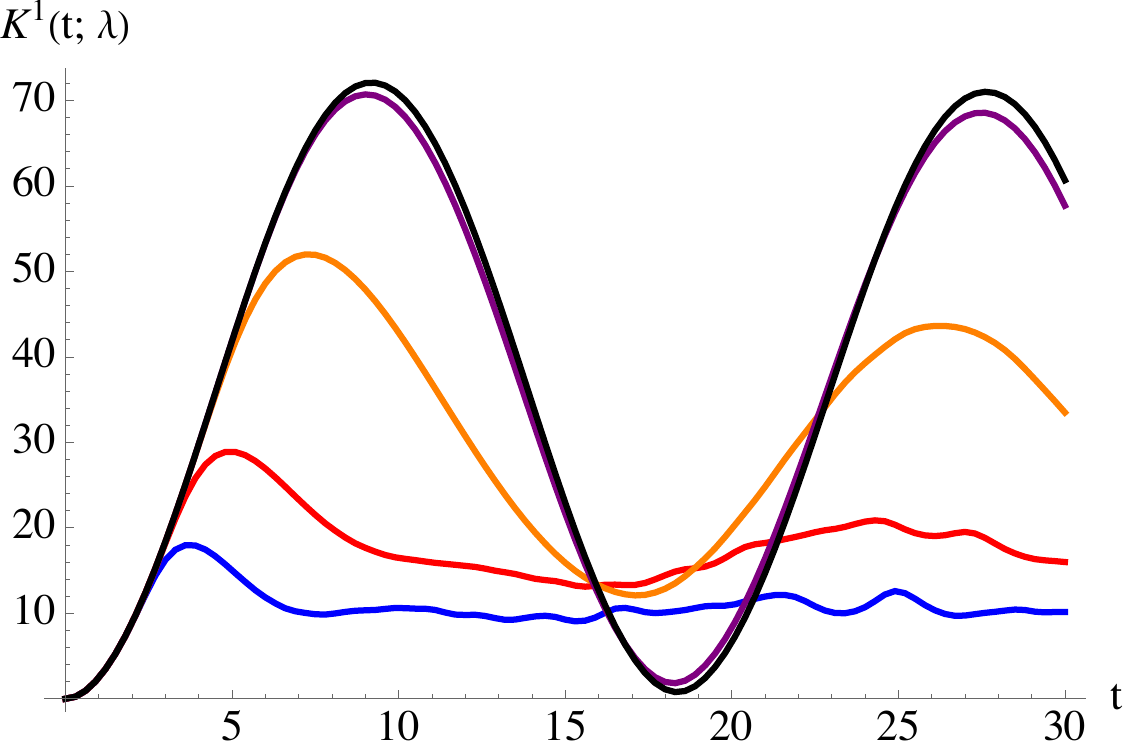}
        \caption{$\l=2/3$}
%        \label{fig:sub1}
    \end{subfigure}
    \qquad
    \begin{subfigure}{0.4\linewidth}
        \centering
        \includegraphics[width=\linewidth]{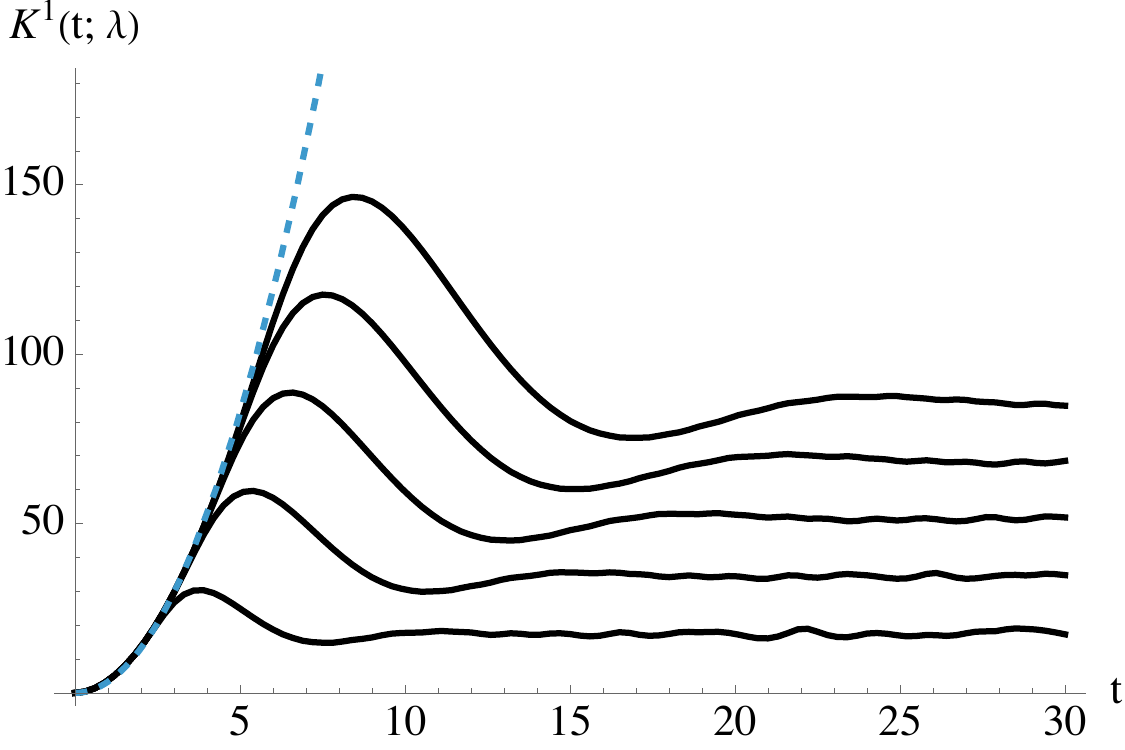}
        \caption{$\l=1$}
%        \label{fig:sub2}
    \end{subfigure}

    \caption{Numerical analysis of fermionic complexity $K^{1}(t; \l)$.  
    The left panel is for $\l=2/3$, a generic value smaller than 1. Blue, red, orange, purple, and black lines correspond to $\K=30,50,100,200,250$.
    Further increasing $\K$ does not change complexity appreciably in this temporal window. 
    The final complexity oscillates. Values $\l>1$ are similar. In the right panel, we show results for $\l=1$. Curves from bottom to top
    correspond to $\K=50,100,150,200,250$. At the critical coupling, we see that complexity saturates for any $\K$ and an enveloping curve emerges. It  grows 
    $\sim t^{2}$ (the dashed blue line  is $3.3\, t^{2}$).}
    \label{fig:VW-K1}
\end{figure}

A similar analysis can be performed in the $F=1$ sector and is illustrated in Fig.~\ref{fig:VW-K1}. 
In both sectors, one sees that for $\l\neq 1$ complexity stabilizes for large $\K$ to an oscillating function. The behaviour 
at the critical point $\l=1$ is quite different. For any $\K$, complexity saturates for large time at a plateau value which increases with $\K$.
The envelope of the curves shows that in this case the asymptotic $\K\to\infty$ complexity is quadratic in time with a ratio of fermionic to bosonic
amplitudes roughly equal to 3. The quadratic rate is replaced by the law $K^{F}_{M}(t)\sim C^{F}_{M}\, t^{2M}$ in the case of the higher degree complexity.

In the following section we introduce a model related to the Veneziano--Wosiek model by a rank-1 deformation. 
This model can be studied analytically and will be shown to exhibit features very close to those of the original model.

\section{The rank-1 modified model}
\la{sec:mod-model}

In this section we introduce a modified model which is obtained by a rank-1 deformation of the $F=0$ Veneziano--Wosiek Hamiltonian, 
first discussed in \cite{DePietri:2006hi}. We will show that it can be extended to the $F=1$ sector in a supersymmetric way.
The resulting modified Hamiltonian $\wt H$ may be treated fully analytically. This is straightforward in the bosonic sector
where the model is in a known class with dynamical $\mk{sl}(2, \mathbb R)$ symmetry, while the fermionic extension is 
non-trivial. The Krylov complexity in the solvable model will be shown to share many qualitative features with that of the original Veneziano--Wosiek model.

\subsection{The modified $\mk{sl}(2,\mathbb R)$ Hamiltonian in $F=0$ sector}

The matrix $H$ in (\ref{3.7}) acts on $F=0$ states $\{\ket{0,n}\}_{n\ge 1}$ as
\be
\la{5.1}
(H\psi)_{n} = [1+\l(1-\delta_{n,1})]\, n \psi_{n}+\sql \sqrt{n(n-1)}\psi_{n-1}+\sql\sqrt{n(n+1)}\psi_{n+1}.
\ee
The modified Hamiltonian corresponds to a  rank 1 perturbation, \ie a change in the $(1,1)$ matrix  element, and reads
\be
\la{5.2}
(\wt H\psi)_{n} = (1+\l)\, n \psi_{n}+\sql \sqrt{n(n-1)}\psi_{n-1}+\sql\sqrt{n(n+1)}\psi_{n+1}.
\ee
It obeys the important exact relation 
\be
\la{5.3}
(\l\wt H_{1/\l}\psi)_{n} = 
(\wt H_{\l}\psi)_{n},
\ee
that implies a manifest $\l\to \l^{-1}$ duality.
We may change variables $\psi_{n}\to \phi_{n}$ by setting 
\be
\la{5.4}
\psi_{n}= (-\sql)^{n}\sqrt n\, \phi_{n}.
\ee
%We have 
%\ba
%(-1)^{n} & (\sql)^{n}\sqrt n (\wt H\phi)_{n} =
% (1+\l)\, n (-1)^{n}(\sql)^{n}\sqrt n \phi_{n}+\sql \sqrt{n(n-1)}(-1)^{n-1}(\sql)^{n-1}\sqrt{n-1}\phi_{n-1}\lp
% +\sql\sqrt{n(n+1)}(-1)^{n+1}(\sql)^{n+1}\sqrt {n+1}\phi_{n+1},
%\ea
The associated action of $\wt H$ on the vector $\{\phi_{n}\}$ is then  
\ba
\la{5.6}
(\wt H\phi)_{n} =
 (1+\l)\, n\phi_{n}-(n-1)\phi_{n-1}-\l(n+1)\phi_{n+1}.
\ea
The Lanczos coefficients are thus 
\be
\la{5.13}
\sfa_{n} = (1+\l)\, n, \qquad \sfb_{n} = \sql\sqrt{n(n-1)}, \qquad n\ge 1.
 \ee
They can be compared with 
Lanczos coefficients of the model with $\mk{sl}(2,\mathbb R)$ symmetry generated by $L_{0}, L_{\pm 1}$
\cite{Caputa:2021sib,Balasubramanian:2022tpr}, see in particular Appendix A of  \cite{Caputa:2024sux}. 
In that framework, one considers the Lanczos chain built by starting from  the highest state $\ket{h}$ with $L_{0}\ket{h}=h\ket{h}$ that corresponds here to the state
$\ket{0; 1}$. The Hamiltonian
has the form $H = \gamma L_{0}+\alpha(L_{1}+L_{-1})$ and we may identify (\ref{5.13}) with 
the $\mk{sl}(2,\mathbb R)$ parameters
\be
h=1, \qquad \gamma=1+\l, \qquad \alpha=\sql.
\ee
%Notice in particular that the survival amplitude is then $\psi_{0}^{*}(t)$ where $\psi_{n}(t)$ is in Eq.~(44) in \cite{Caputa:2024sux}.
The associated spectral measure is the one for Meixner-Pollaczek polynomials, see \cite{Muck:2022xfc}.
For our purposes, it is convenient to recall a few facts about the spectral properties of (\ref{5.13}).
The spectrum is discrete
\be
\la{5.7}
\wt E_{p}=(1-\l)p, \qquad p=1,2,\dots,
\ee
with eigenvectors that for $\l<1$ are Meixner polynomials
\ba
\phi^{(p)}_{n} &= (1-\l)\,\sqrt{p}\,\l^{-p/2}\, F(-(p-1), n+1, 2, 1-\l),\\
& \sum_{n=1}n \l^{n}\phi^{(p)}_{n}\phi^{(q)}_{n} = \delta^{pq}.
\ea
In terms of $\psi_{n}^{(p)}$, \cf (\ref{5.4}), these relations read
\ba
\la{5.10}
 \psi^{(p)}_{n} &= (-1)^{n}\sqrt{np} (1-\l)\,\l^{(n-p)/2}\, F(-(p-1), n+1, 2, 1-\l),\\
 & \sum_{n=1}\psi^{(p)}_{n}\psi^{(q)}_{n} = \delta^{pq}.
 \ea
 The above sums  converge for $0<\l<1$. For $\l>1$ we should use instead
 \be
 \phi_{n}^{(p)} = (1-\l)\sqrt{p}\l^{-n-\frac{p}{2}} F(-(p-1), 1-n, 2, 1-\l), 
 \ee
 with spectrum $\wt E_{p} = (\l-1)p$, $p=1, 2, \dots$. 
 In the limit $\l\to 1$, the eigenvalues vanish linearly in $\l-1$, which is considerably simpler than the scaling behaviour found in (\ref{3.27}).

 \subsubsection{Zero modes}
 
 Notice that we do not have non-trivial normalizable zero modes for any $\l$. The change of variable (\ref{5.4})
 is well defined for $n\ge 1$ and the modified problem (\ref{5.2}) has only the trivial zero mode $\ket{0}$ for any $\l$. Notice that looking for a second zero energy state 
 solving (\ref{5.1}) with $\psi_{1}=1$ gives the Veneziano--Wosiek solution, \cf (\ref{3.20}) up to an overall normalization,
 \be
 \psi_{n} = (-1)^{n+1}\frac{1}{\sqrt{n}}\l^{\frac{1-n}{2}},
 \ee
which is normalizable for $\l>1$. Instead, solving for a zero energy state in (\ref{5.2}) we find 
\be
\psi_{n} = (-1)^{n+1}\frac{1}{\sqrt{n}} \frac{\l^{n/2}-\l^{-n/2}}{\l^{1/2}-\l^{-1/2}},
\ee
which is not normalizable for any $\l$ (and is of course symmetric under $\l\to 1/\l$). 
The spectrum of the original Veneziano--Wosiek model can be recovered by exploiting the properties of rank-1 deformations
as discussed in \cite{DePietri:2006hi}.

\subsubsection{Spectrum at $\lambda=1$ and truncated finite-dimensional space}
 
Near $\lambda=1$, the Veneziano–Wosiek model exhibits the scaling (\ref{3.27}), while the modified model shows the simple linear behavior $E_{n}\sim (\lambda-1)$, cf.\ (\ref{5.7}). 
It is interesting to investigate what happens exactly at $\lambda=1$ by truncating the state space with a boson number cutoff. 
In the modified model this analysis is particularly simple and transparent.

Let us look for an eigenvector of (\ref{5.6}) with eigenvalue $E$ at $\lambda=1$:
\be
\la{5.15}
(\widetilde H\phi)_{n}=2n\phi_{n}-(n-1)\phi_{n-1}-(n+1)\phi_{n+1} = E \phi_{n}.
\ee
The associated Laguerre polynomials $L_{n}^{(1)}(E)$ satisfy the recursion relation
\be
(n+1)L_{n+1}^{(1)}(E) = (2n+2-E)L_{n}^{(1)}(E)-(n+1)L_{n-1}^{(1)}(E).
\ee
Shifting the indices gives
\be
nL_{n}^{(1)}(E) = (2n-E)L_{n-1}^{(1)}(E)-nL_{n-2}^{(1)}(E),
\ee
which can be rewritten as
\be
2L_{n-1}^{(1)}(E)-L_{n-2}^{(1)}(E)-L_{n}^{(1)}(E) = \frac{E}{n}L_{n-1}^{(1)}(E).
\ee
Therefore, setting
\be
\la{5.19}
\phi_{n} = \frac{1}{n}L_{n-1}^{(1)}(E),
\ee
we obtain
\be
(\widetilde H\phi)_{n}
=2L_{n-1}^{(1)}(E)-L_{n-2}^{(1)}(E)-L_{n}^{(1)}(E)
= \frac{E}{n}L_{n-1}^{(1)}(E)
= E\phi_{n}.
\ee
Thus (\ref{5.19}) solves the recursion (\ref{5.15}) with the boundary condition $\phi_{1}=1$.

Introducing a cutoff $\K$ on the boson number, the spectrum $\{E_{k}(\K)\}_{k=1,2,\dots}$ is determined by the condition
\be
\phi_{\K}(E_{k}(\K))=0,
\ee
where $\K$ plays a role analogous to the dimension of the Lanczos basis.

By using the known properties of associated Laguerre polynomials \footnote{See e.g.\ {\tt https://dlmf.nist.gov/18.16}.}, one obtains for large $\K$ the estimate
\be
E_{k}(\K) \simeq \frac{1}{4\K}j_{1,k}^{2},
\ee
where $j_{1,k}$ denotes the $k$-th positive zero of the Bessel function $J_{1}(x)$. This analytic result confirms that the spectrum collapses to a continuum in the limit $\K\to \infty$.
As an illustration, for $\K=10$ and $\K=100$ the first three eigenvalues are
\be
\def\arraystretch{1.3}
\begin{array}{cccc}
\toprule
 \K=10  \\
\midrule
E_{k}(\K) & 0.334528676325 & 1.12825335588 & 2.39586992475  \\
\frac{1}{4\K}j^{2}_{1,k} & 0.367049266053 & 1.23046140804 & 2.58748634738  \\
\midrule
 \K=100  \\
\midrule
E_{k}(\K) & 0.0363426012426 & 0.121840110669 & 0.256240943267  \\
\frac{1}{4\K}j^{2}_{1,k} & 0.0367049266053 & 0.123046140804 & 0.258748634738  \\
\bottomrule
\end{array}
\ee

\subsubsection{Number -- Energy duality}

Notice the further relation
\be
\la{5.24}
\mmm{\psi^{(p)}}{\wh N}{\psi^{(q)}} = \sum_{n=1}n\,\psi^{(p)}_{n}\psi^{(q)}_{n} = \frac{1}{1-\l}\bigg[p(1+\l)\delta_{p,q}-\sql\sqrt{q(q+1)}\delta_{p, q+1}-\sql\sqrt{q(q-1)}\delta_{p,q-1}\bigg].
\ee
This means that the operator $\wh N$ is same as $\wt H$ in the eigenvector basis, up to a further $\psi_{n}\to (-1)^{n}\psi_{n}$ transformation,
consistently with  (\ref{5.7}).

\subsection{Supersymmetry and solvable model in $F=1$ sector}
 
Consider states $\ket{0,n}$ with $n\ge 1$ and $\ket{1,n}$ with $n\ge 0$. 
Let us introduce the nilpotent supercharge $\wt Q$ with action 
\bea
\wt Q \ket{0, n} &= 0,\\
\wt Q\ket{1,n}  &=\sqrt\frac{1-\l^{n+2}}{1-\l^{n+1}}\sqrt{n+1}\ket{0, n+1}+\sql\sqrt\frac{1-\l^{n+1}}{1-\l^{n+2}}\sqrt{n+2}\ket{0,n+2}, \\ 
\wt Q^{\dag} \ket{1, n} &= 0, \\
\wt Q^{\dag}\ket{0,n} &= \sqrt\frac{1-\l^{n+1}}{1-\l^{n}}\sqrt{n}\ket{1, n-1}+\sql\sqrt\frac{1-\l^{n-1}}{1-\l^{n}}\sqrt{n}\ket{1,n-2},
\eea
where we notice the fermionic duality
\be
\wt Q_{\l} = \sql\, \wt Q_{1/\l}.
\ee
In the $F=0$ sector we can write
\be
\la{5.27}
\wt H^{F=0} = \{\wt Q, \wt Q^{\dag}\}.
\ee
This follows from a straightforward computation
\ba
(\wt Q\wt Q^{\dag} & + \wt Q^{\dag}\wt Q)\ket{0, n} = \wt Q\wt Q^{\dag}\ket{0,n} = 
\wt Q\bigg(\sqrt\frac{1-\l^{n+1}}{1-\l^{n}}\sqrt{n}\ket{1, n-1}+\sql\sqrt\frac{1-\l^{n-1}}{1-\l^{n}}\sqrt{n}\ket{1,n-2}\bigg)\lp
= \sqrt\frac{1-\l^{n+1}}{1-\l^{n}}\sqrt{n}\bigg(\sqrt\frac{1-\l^{n+1}}{1-\l^{n}}\sqrt{n}\ket{0, n}+\sql\sqrt\frac{1-\l^{n}}{1-\l^{n+1}}\sqrt{n+1}\ket{0,n+1}\bigg)\lp
+\sql\sqrt\frac{1-\l^{n-1}}{1-\l^{n}}\sqrt{n}\bigg(\sqrt\frac{1-\l^{n}}{1-\l^{n-1}}\sqrt{n-1}\ket{0, n-1}+\sql\sqrt\frac{1-\l^{n-1}}{1-\l^{n}}\sqrt{n}\ket{0,n}\bigg)\lp
= n(1+\l)\ket{0,n}+\sql\sqrt{n(n+1)}\ket{0, n+1}+\sql\sqrt{n(n-1)}\ket{0,n-1},
\ea
which agrees with (\ref{5.2}). 
Note that while $\wt H$ is close to $H$,
the supercharge $\wt Q$ is definitely more complicated than (\ref{3.13}).
Still, we may define the partner Hamiltonian $\wt H^{F=1}$ by imposing the same relation (\ref{5.27}) in $F=1$ sector.
This gives
\ba
\wt H^{F=1}\ket{1, n} &= \{\wt Q, \wt Q^{\dag}\}\ket{1,n} = \bigg[\frac{(2+n)\l(\l^{n+1}-1)}{\l^{n+2}-1}+\frac{(n+1)(\l^{n+2}-1)}{\l^{n+1}-1}\bigg]\ket{1,n}\lp
-(n+2)\sql\frac{\sqrt{(\l^{n+3}-1)(\l^{n+1}-1)}}{\l^{n+2}-1}\ket{1,n+1}
-(n+1)\sql\frac{\sqrt{(\l^{n+2}-1)(\l^{n}-1)}}{\l^{n+1}-1}\ket{1,n-1}.
\ea
In the limit $\l\to 1$ we get 
\ba
\wt H^{F=1}_{\l=1}\ket{1, n} &= (2n+3)\ket{1,n}+\sqrt{(n+1)(n+3)}\ket{1,n+1}+\sqrt{n(n+2)}\ket{1, n-1}.
\ea
The eigenvectors of $\wt H^{F=1}$ are obtained by acting with $\wt Q^{\dag}$ on $F=0$ eigenvectors
\be
\wt H^{F=1}\wt Q^{\dag}\ket{\psi^{(p)}} = \wt Q^{\dag}\wt Q \wt Q^{\dag}\ket{\psi^{(p)}} = E_{p}\wt Q^{\dag}\ket{\psi^{(p)}},
\ee
and the  orthonormal ones are
\be
\la{5.32}
\ket{\chi^{(p)}} = \frac{1}{\sqrt{(1-\l)p}}\wt Q^{\dag}\ket{\psi^{(p)}}.
\ee

\section{Analytical  Krylov complexity in the modified model}
\la{sec:mod-analytical}

Due to the better analytical structure of $\wt H$, compared to $H$, we may compute analytically the (higher degree) Krylov
complexity $K_{M}^{F}(t; \l)$ for generic planar coupling $\l$. In the bosonic sector, one can rely
on the results and methods in \cite{Caputa:2021sib,Balasubramanian:2022tpr}, based on $\mk{sl}(2,\mathbb R)$ symmetry. 
However, we prefer to present here a slightly different computation that will be easily adaptable to the fermionic sector.

\subsection{Bosonic sector $F=0$}
\la{sec:mod-F0}

Introducing resolutions of the identity, we have
\ba
K_{M}^{F=0}(t; \l) &= \mmm{1}{e^{-itH}(\wh N-1)^{M} e^{it H}}{1} = \sum_{p,q=1}^{\infty}e^{-it(1-\l)(p-q)}\braket{1}{\psi^{(p)}}\,
\mmm{\psi^{(p)}}{(\wh N-1)^{M}}{\psi^{(q)}}\, \braket{\psi^{(q)}}{1}\lp
= \sum_{p,q,n=1}^{\infty}e^{-it(1-\l)(p-q)}\braket{1}{\psi^{(p)}}\,\braket{\psi^{(p)}}{n}(n-1)^{M}\braket{n}{\psi^{(q)}}\, \braket{\psi^{(q)}}{1}\lp
= \sum_{n=1}^{\infty}(n-1)^{M}|\mc B_{n}(t; \l)|^{2},
\ea
where we introduced the quantity
\ba
\la{6.2}
\mc B_{n}(t; \l) = \sum_{p=1}^{\infty}e^{-it(1-\l)p}\psi^{(p)}_{1}\psi^{(p)}_{n}.
\ea
This sum can be computed in closed form, see Appendix \ref{app:sum} for full details,  and one obtains 
\ba
\la{6.3}
|\mc B_{n}(t; \l)|^{2} = 4^{n-1}\,n\,(1-\l)^{4}\,\l^{n-1}\frac{\sin^{2(n-1)}(\frac{1}{2}(1-\l)t)}{[1+\l^{2}-2\l\cos((1-\l)t)]^{n+1}}.
\ea
Then
\ba
K_{M}^{F=0}(t; \l) &= \frac{(1-\l)^{4}}{4\l\, [(1+\l^{2}-2\l\cos((1-\l)t))\sin^{2}(\frac{1}{2}(1-\l)t)]}\sum_{n=1}^{\infty}(n-1)^{M}n X^{n},\\
X &= \frac{4\l\sin^{2}(\frac{1}{2}(1-\l)t)}{1+\l^{2}-2\l\cos((1-\l)t)}.
\ea
Notice that this can be written by expressing the infinite sum in terms of polylogarithms with negative integer index
\ba
\la{6.6}
K_{M}^{F=0}(t; \l) &= \frac{(1-\l)^{4}}{(1+\l^{2}-2\l\cos((1-\l)t))^{2}}\times \lp
\bigg[\LI_{-(M+1)}\bigg(\frac{4\l\sin^{2}(\frac{1}{2}(1-\l)t)}{1+\l^{2}-2\l\cos((1-\l)t)}\bigg)
+\LI_{-M}\bigg(\frac{4\l\sin^{2}(\frac{1}{2}(1-\l)t)}{1+\l^{2}-2\l\cos((1-\l)t)}\bigg)\bigg].
\ea
This is a closed formula that may be evaluated readily for any desired $M=1, 2, \dots$. We find
\ba
K^{F=0}_{1}(t; \l) &= \frac{8\l}{(1-\l)^{2}}\sin^{2}\bigg(\frac{1}{2}(1-\l)\, t\bigg), \\
K^{F=0}_{2}(t; \l) &= \frac{8\l}{(1-\l)^{2}}\sin^{2}(\tfrac{1}{2}(1-\l)\,t)+\frac{96\l^{2}}{(1-\l)^{4}}\sin^{4}(\tfrac{1}{2}(1-\l)\,t), \\
K^{F=0}_{3}(t; \l) &= \frac{8\l}{(1-\l)^{2}}\sin^{2}(\tfrac{1}{2}(1-\l)\,t)+\frac{288\l^{2}}{(1-\l)^{4}}\sin^{4}(\tfrac{1}{2}(1-\l)\,t)+\frac{1536\l^{3}}{(1-\l)^{6}}\sin^{6}(\tfrac{1}{2}(1-\l)\,t), 
\ea
and so on. These are exact expressions valid for any $\l$. In particular, the standard Krylov complexity $M=1$ has a simple squared sinus form and oscillates. This behaviour is similar 
to what happens in the Veneziano--Wosiek model. Notice that from the representation (\ref{6.6}) we can check the exact weak-strong duality, \cf (\ref{5.3}),
\be
\la{6.10}
K^{F=0}(t; \l) = K^{F=0}(\l t, 1/\l).
\ee

\paragraph{$\l=1$ limit}

We can compute analytically the $\l=1$ limit from specialization of (\ref{6.6})
\ba
K_{M}^{F=0}(t) &= \frac{1}{(1+t^{2})^{2}}
\bigg[\LI_{-(M+1)}\bigg(\frac{t^{2}}{1+t^{2}}\bigg)
+\LI_{-M}\bigg(\frac{t^{2}}{1+t^{2}}\bigg)\bigg].
\ea
Explicit values are
\bea
\la{6.12}
K_{1}^{F=0}(t) &= 2t^{2}, \\
K_{2}^{F=0}(t) &= 2t^{2}(1+3t^{2}), \\
K_{3}^{F=0}(t) &= 2t^{2}(1+9t^{2}+12t^{4}), \\
K_{4}^{F=0}(t) &= 2t^{2}(1+21t^{2}+72t^{4}+60t^{6}).
\eea
Again, for $M=1$ we get an exact $t^{2}$ growth as suggested by numerics in the Veneziano--Wosiek model. We can get an explicit recursion for the polynomials in (\ref{6.12}).
To this aim, let us consider the function 
\be
\la{6.13}
S_{M} = \LI_{-M-1}(x)+\LI_{-M}(x),
\ee
that obeys \footnote{This is valid for both terms in (\ref{6.13}) since $x\frac{d}{dx}\LI_{-M}(x) = x\frac{d}{dx}\sum_{n=1}^{\infty}n^{M}x^{n} = \LI_{-(M+1)}(x).$}
\be
S_{M} = x\frac{d}{dx}S_{M-1}(x).
\ee
Thus,
\be
(1+t^{2})^{2}K_{M}^{F=0} = \frac{t^{2}}{1+t^{2}}\frac{d}{d\frac{t^{2}}{1+t^{2}}}[(1+t^{2})^{2}K_{M-1}^{F=0}],
\ee
that implies
\be
\la{6.16}
K^{F=0}_{M}(t) = \bigg[2t^{2}+\frac{1}{2}t(1+t^{2})\frac{d}{dt}\bigg]K_{M-1}^{F=0}(t).
\ee
Hence, if we write
\be
K^{F=0}_{M}(t) = 2t^{2}P_{M}(t), \qquad P_{1}(t) = 1,  
\ee
we get the following recursion for $P_{M}(t)$ 
\be
\la{6.18}
P_{M}(t) = \bigg[1+3t^{2}+\frac{1}{2}t(1+t^{2})\frac{d}{dt}\bigg]\,P_{M-1}(t),
\ee
from which it follows that $P_{M}(t)$ is a polynomial in $t^{2}$ of degree $M-1$ and positive integer coefficients, as in (\ref{6.12}). 
The leading terms of $P_{M}(t)$ at large $t$ are
\be
\la{6.19}
P_{M}(t) = \frac{1}{2}(M+1)!\, t^{2(M-1)}+\frac{1}{4}M!\, M(M-1)\, t^{2(M-2)}+\cdots
\ee
A fully explicit determination of $P_{M}(t)$ is discussed in Appendix \ref{app:poly}, with the result
\be
K_{M}^{F=0}(t) = \sum_{k=1}^{M}(k+1)!\, S_{2}(M,k)\, t^{2k},
\ee
where $S_{2}(M,k)$ are Stirling numbers of second kind, cf.  (\ref{C.8}).

\subsection{Fermionic sector $F=1$}

In the $F=1$ sector we recall that $\ket{s_{n}} = \ket{1; n}$ and using a resolution of the identity based 
on the fermionic eigenstates (\ref{5.32}) we get  
\ba
K^{F=1}_{M}(t; \l) &= \mmm{1,0}{e^{-itH}\,\wh N^{M}\, e^{it H}}{1,0}  \lp
= \sum_{p,q=1}^{\infty}e^{-it(1-\l)(p-q)}\braket{1,0}{\chi^{(p)}}\,
\mmm{\chi^{(p)}}{\wh N^{M} }{\chi^{(q)}}\, \braket{\chi^{(q)}}{1,0}.
\ea
The scalar products are 
\ba
\braket{1,0}{\chi^{(p)}} &= \frac{1}{\sqrt{(1-\l)p}}\mmm{1,0}{\wt Q^{\dag}}{\psi^{(p)}} = \frac{1}{\sqrt{(1-\l)p}}\bigg(\sqrt{1+\l}\bra{0,1}+\sqrt\frac{2\l}{1+\l}\bra{0,2}\bigg)\ket{\psi^{(p)}}\lp
= -p\frac{(1-\l)^{2}}{\sqrt{\l(1-\l^2)}}\l^{p/2}.
\ea
The $\wh N^{M}$ matrix element may be written  
\ba
& \mmm{\chi^{(p)}}{\wh N^{M} }{\chi^{(q)}} = \frac{1}{\sqrt{(1-\l)p}}\frac{1}{\sqrt{(1-\l)q}}\mmm{\psi^{(p)}}{\wt Q\, \wh N^{M}\,\wt Q^{\dag}}{\psi^{(q)}} \lp
= \frac{1}{1-\l}\frac{1}{\sqrt{pq}}\sum_{n=0}^{\infty}n^{M}\,\mmm{\psi^{(p)}}{\wt Q}{1,n}  \mmm{\psi^{(q)}}{\wt Q}{1,n}.
\ea
Using
\ba
 \mmm{\psi^{(q)}}{\wt Q}{1,n} &=\sqrt\frac{1-\l^{n+2}}{1-\l^{n+1}}\sqrt{n+1}\,\psi^{(q)}_{n+1}+\sql\sqrt\frac{1-\l^{n+1}}{1-\l^{n+2}}\sqrt{n+2}\,\psi^{(q)}_{n+2},
 \ea
we get finally
\ba
\la{6.25}
K^{F=1}_{M}(t; \l) &= \sum_{n=1}^{\infty}n^{M}\, |\mc F_{n}(t; \l)|^{2},
\ea
with
\ba
\mc F_{n}(t; \l) &= \frac{1-\l}{\sqrt{\l(1+\l)}}\sum_{p=1}^{\infty} e^{-i(1-\l)pt}\,\sqrt{p}\,\l^{p/2}\sqrt\frac{1-\l^{n+2}}{1-\l^{n+1}}
\bigg(\sqrt{n+1}\psi^{(p)}_{n+1}+\sql\frac{1-\l^{n+1}}{1-\l^{n+2}}\sqrt{n+2}\psi^{(p)}_{n+2}\bigg).
\ea
For each $n$ we can compute the sum over $p$ and we find
\ba
\la{6.27}
|\mc F_{n}(t; \l)|^{2} &= 2^{2n}\frac{(1-\l)^{6}}{(1+\l)^{2}}\,\l^{n}\frac{\sin^{2n}\mu\, \,(P_{1,n}(\l)^{2}\cos^{2}\mu+P_{2,n}(\l)^{2}\sin^{2}\mu)}{P_{3,n}(\l)\, (1+\l^{2}-2\l\cos(2\mu))^{n+3}},
\quad \mu\equiv\frac{1}{2}(1-\l)\,t,
\ea
where (the finite sums in $P_{2}$ and $P_{3}$ are of course trivial)
\ba
& P_{1,n}(\l) = (n+1)\frac{1-\l^{n+2}}{1-\l}, \qquad
 \begin{cases}
P_{2,2n}(\l) = \sum_{m=0}^{n}(2m+1)\l^{n-m}(1-\l^{2m+1}), \\
P_{2,2n+1}(\l) = \sum_{m=1}^{n+1}2m\l^{n-m+1}(1-\l^{2m}), 
\end{cases} \\
& P_{3,n}(\l) = \sum_{m=0}^{N_{n}}\l^{2m}\,\sum_{m=0}^{2N_{n+1}}\l^{m}, \qquad N_{n}= \frac{1}{4}(2n-1+(-1)^{n}) = 0,1,1,2,2,3,3,\dots.
\ea
From these expressions one can check the exact weak-strong duality
\be
K^{F=1}(t; \l) = K^{F=1}(\l t; 1/\l),
\ee
which is same as in bosonic sector, \cf (\ref{6.10}). The explicit form of $|\mc F_{n}(t; \l)|^{2}$ for the first values of $n$ are ($\mu\equiv(1-\l)t$)
\bea
|\mc F_{1}(t; \l)|^{2} &= 2^{2}(1-\l)^{6}\l\sin^{2}\frac{\mu}{2}\frac{4(1+\l+\l^{2})^{2}\cos^{2}\frac{\mu}{2}+(2-2\l^{2})^{2}\sin^{2}\frac{\mu}{2}}
{(1+\l)^{2}(1+\l+\l^{2})(1+\l^{2}-2\l\cos\mu)^{4}},\\
%%%
|\mc F_{2}(t; \l)|^{2} &= 2^{4}(1-\l)^{6}\l^{2}\sin^{4}\frac{\mu}{2}\frac{9(1+\l+\l^{2}+\l^{3})^{2}\cos^{2}\frac{\mu}{2}+(3+\l-\l^{2}-3\l^{3})^{2}\sin^{2}\frac{\mu}{2}}
{(1+\l)^{2}(1+\l^{2})(1+\l+\l^{2})(1+\l^{2}-2\l\cos\mu)^{5}},\\
%%%
|\mc F_{3}(t; \l)|^{2} &= 2^{6}(1-\l)^{6}\l^{3}\sin^{6}\frac{\mu}{2}\frac{16(1+\l+\l^{2}+\l^{3}+\l^{4})^{2}\cos^{2}\frac{\mu}{2}+(4+2\l-2\l^{3}-4\l^{4})^{2}\sin^{2}\frac{\mu}{2}}
{(1+\l)^{2}(1+\l^{2})(1+\l+\l^{2}+\l^{3}+\l^{4})(1+\l^{2}-2\l\cos\mu)^{6}},\\
\eea
and so on.
At small $t$, we have $|\mc F_{n}(t; \l)|^{2} = \mc O(t^{2n})$ and each term in the small $t$ expansion receives contributions
from a finite number of values of $n$. We thus get the exact expansions
\ba
\la{6.32}
K^{F=1}_{M=1}(t; \l) &=\frac{4\l(1+\l+\l^{2})}{(1+\l)^{2}}\,t^{2}-\frac{\l(1-\l)^{4}(1+4\l+\l^{2})}{3(1+\l)^{2}(1+\l+\l^{2})}\,t^{4}\lp
+\frac{\l(1-\l)^{4}(1+2\l-41\l^{2}-356\l^{3}-41\l^{4}+2\l^{5}+\l^{6})}{90(1+\l)^{2}(1+\l^{2})(1+\l+\l^{2})}t^{6}+\cdots, \\
%%%
\la{6.33}
K^{F=1}_{M=2}(t; \l) &=\frac{4 \l  (1+\l +\l ^2) }{(1+\l 
)^2}\,t^{2}+\frac{\l  (-1+54 \l +117 \l ^2+92 \l ^3+117 
\l ^4+54 \l ^5-\l ^6)}{3 (1+\l )^2 (1+\l 
+\l ^2)}\,t^{4}\lp
+\frac{(1-\l )^4 \l  (1-268 \l -1121 
\l ^2-2336 \l ^3-1121 \l ^4-268 \l ^5+\l ^6) 
}{90 (1+\l )^2 (1+\l ^2) (1+\l +\l ^2)}\,t^{6}+\cdots.
\ea
and so on.

\paragraph{Specialization to $\l=1$ limit}

Taking $\l\to 1$ is simple by using the relation 
\be
|\mc F_{n}(t; \l)|^{2} = \frac{1}{2}(n+1)(n+2)\frac{t^{2n}}{(1+t^{2})^{n+3}}+\mc O(\l-1).
\ee
Then,
\ba
K^{F=1}_{M}(t) &= \frac{1}{2(1+t^{2})^{3}} \sum_{n=1}^{\infty}n^{M}(n+1)(n+2)\bigg(\frac{t^{2}}{1+t^{2}}\bigg)^{n}
\ea
Again, we can write this sum in terms of polylogarithms with negative integer index. However, it is convenient to give a recursion for the result. We start from 
\be
S_{M}(x) =  \sum_{n=1}^{\infty}n^{M}(n+1)(n+2)x^{n} = x\frac{d}{dx}S_{M-1}(x), \qquad S_{1}(x) = \frac{6x}{(1-x)^{4}}.
\ee
Then,
\be
2(1+t^{2})^{3}K_{M}^{F=1}(t) = \frac{t^{2}}{1+t^{2}}\frac{d}{d\frac{t^{2}}{1+t^{2}}}[2(1+t^{2})^{3}K_{M-1}^{F=1}(t)],
\ee
that is, \cf (\ref{6.16}),
\be
\la{6.38}
K^{F=1}_{M}(t) = \bigg[3t^{2}+\frac{1}{2}t(1+t^{2})\frac{d}{dt}\bigg]K^{F=1}_{M-1}(t).
\ee
The Ansatz
\be
K^{F=1}_{M}(t) = 3t^{2}Q_{M}(t), \qquad Q_{1}(t)=1,
\ee
gives the simple differential recursion similar to (\ref{6.18})
\be
\la{6.40}
Q_{M}(t) = \bigg[1+4t^{2}+\frac{1}{2}t(1+t^{2})\frac{d}{dt}\bigg]\, Q_{M-1}(t).
\ee
Again, this implies that $Q_{M}(t)$ is a polynomial in $t^{2}$ of degree $M-1$ and positive integer coefficients. 
The leading terms of $Q_{M}$ at large $t$ are
\be
\la{6.41}
Q_{M}(t) = \frac{1}{6}(M+2)!\, t^{2(M-1)}+\frac{1}{12}(M+1)! M(M-1)\, t^{2(M-2)}+\cdots\, .
\ee
A fully explicit determination of $Q_{M}(t)$ follows from the results in Appendix \ref{app:poly} that imply
\be
K^{F=1}_{M}(t) =  \frac{1}{2}\sum_{k=1}^{M}(k+2)!\, S_{2}(M,k)\, t^{2k}.
\ee
The first cases  are 
\bea
\la{6.43}
K^{F=1}_{1}(t) &= 3t^{2}, \\
K^{F=1}_{2}(t) &= 3t^{2}(1+4t^{2}), \\
K^{F=1}_{3}(t) &= 3t^{2}(1+2t^{2})(1+10t^{2}), \\
K^{F=1}_{4}(t) &= 3t^{2}(1+28t^{2}+120t^{4}+120t^{6}),
\eea
and are clearly consistent with the truncation of (\ref{6.32}, \ref{6.33}) for $\l=1$.

\subsection{Numerical analysis with a truncated Lanczos basis}

In this section, we discuss how the exact expressions for $K^{F}_{M}$ derived above are reproduced by numerical evaluation
using a truncated Lanczos basis. The aim is again to show convergence as $\K$ is increased.

\begin{figure}[htbp]
    \centering

    \begin{subfigure}{0.4\linewidth}
        \centering
        \includegraphics[width=\linewidth]{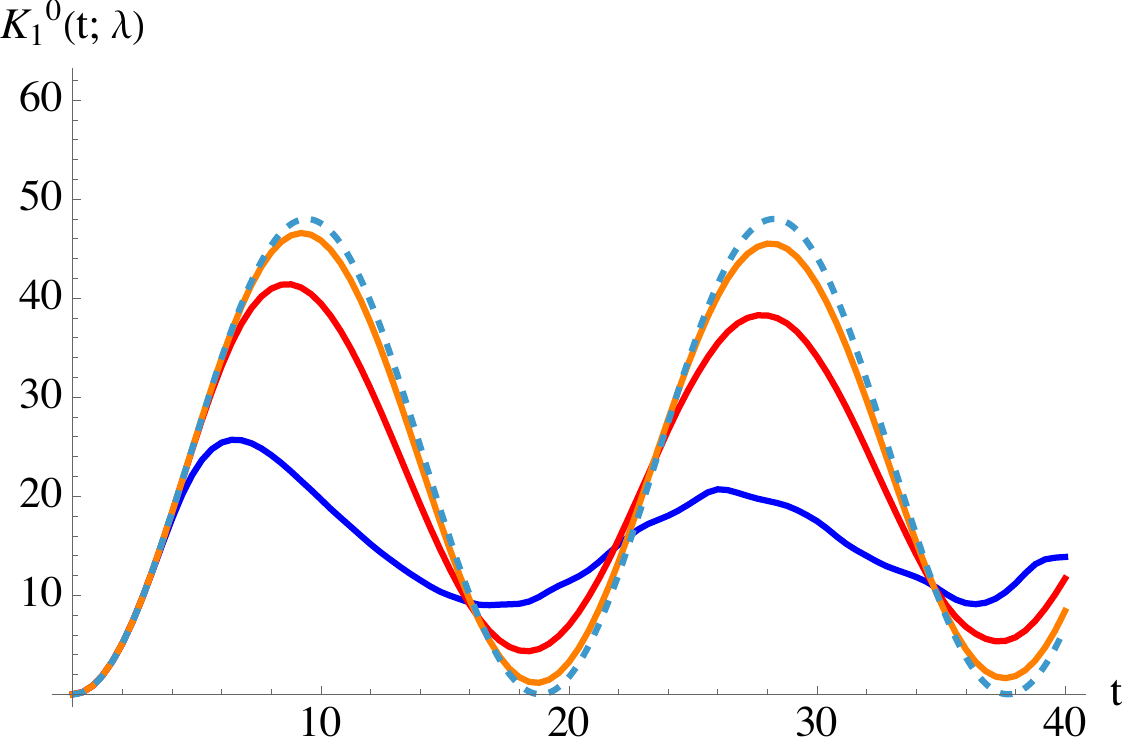}
        \caption{$\l=2/3$}
    \end{subfigure}
    \qquad
    \begin{subfigure}{0.4\linewidth}
        \centering
        \includegraphics[width=\linewidth]{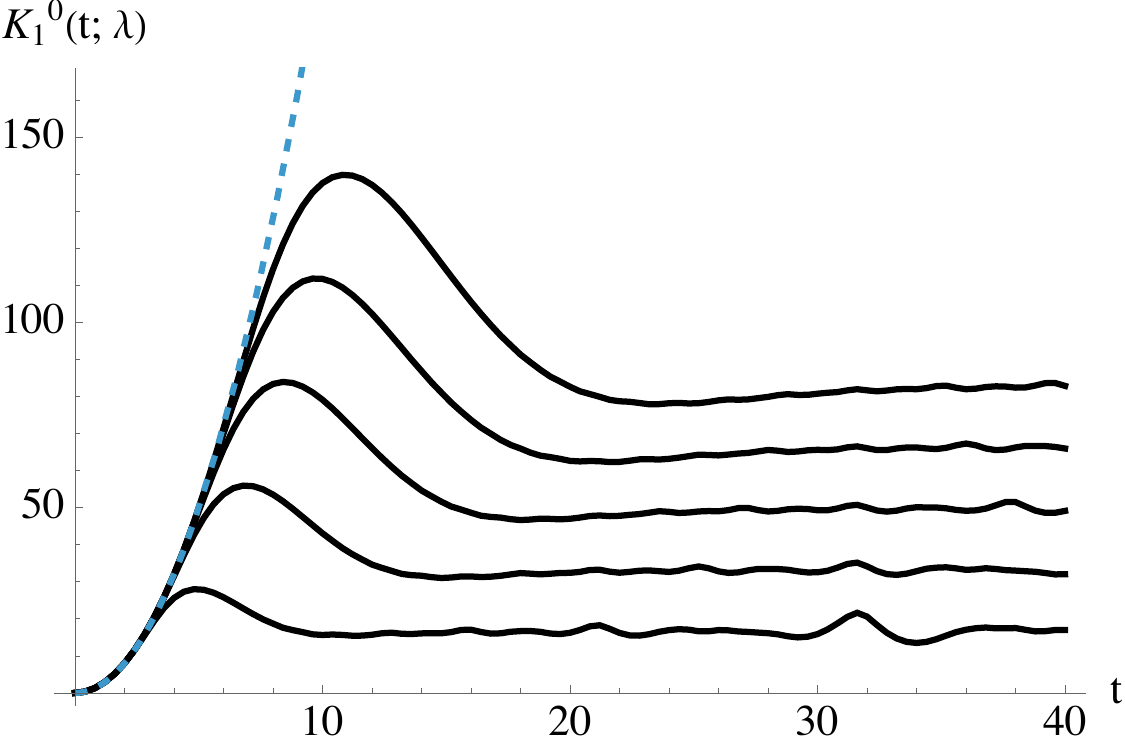}
        \caption{$\l=1$}
    \end{subfigure}

    \caption{
    Analysis of the sector $F=0$ for $M=1$. The curves in panel (a) for $\l=2/3$ correspond to $\K=50,100,150$, while the dashed line represents the exact result (\ref{6.6}).
    The curves in panel (b) for $\l=1$ correspond to $\K=50,100,150,200,250$, and the enveloping dashed line is the first expression in 
    (\ref{6.12}), \ie the quadratic function $2t^{2}$.
    }
    \label{fig:Mod-K01}
\end{figure}

\begin{figure}[htbp]
    \centering

    \begin{subfigure}{0.4\linewidth}
        \centering
        \includegraphics[width=\linewidth]{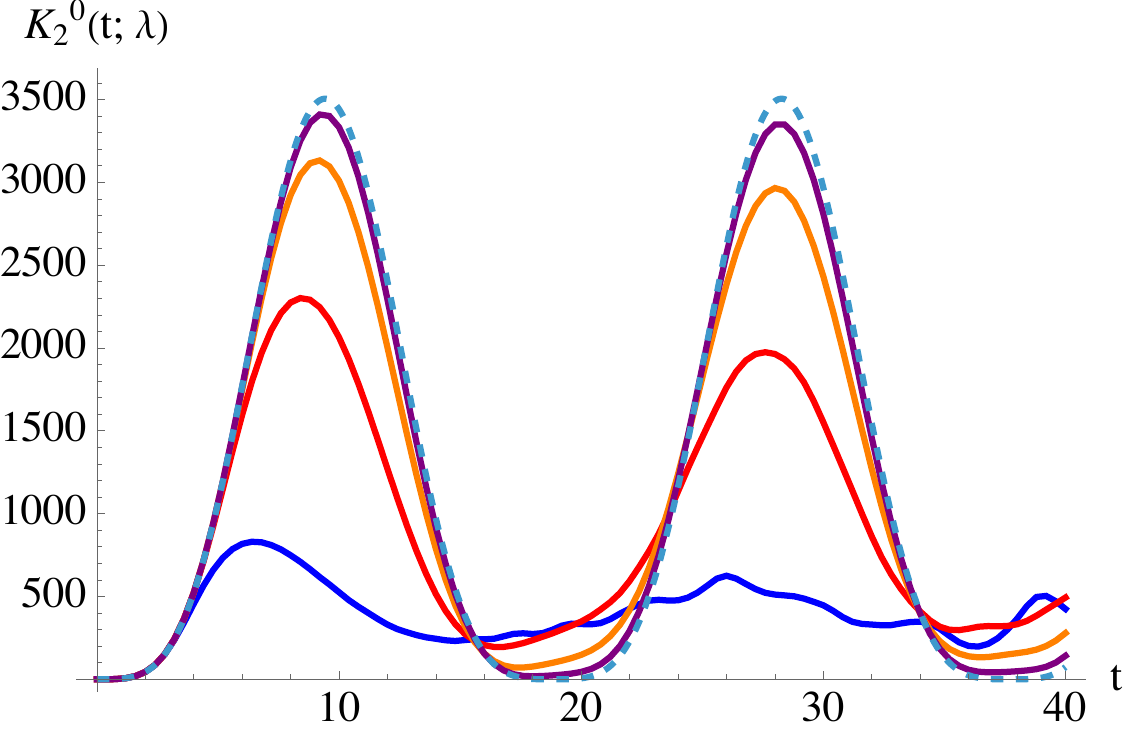}
        \caption{$\l=2/3$}
    \end{subfigure}
    \qquad
    \begin{subfigure}{0.4\linewidth}
        \centering
        \includegraphics[width=\linewidth]{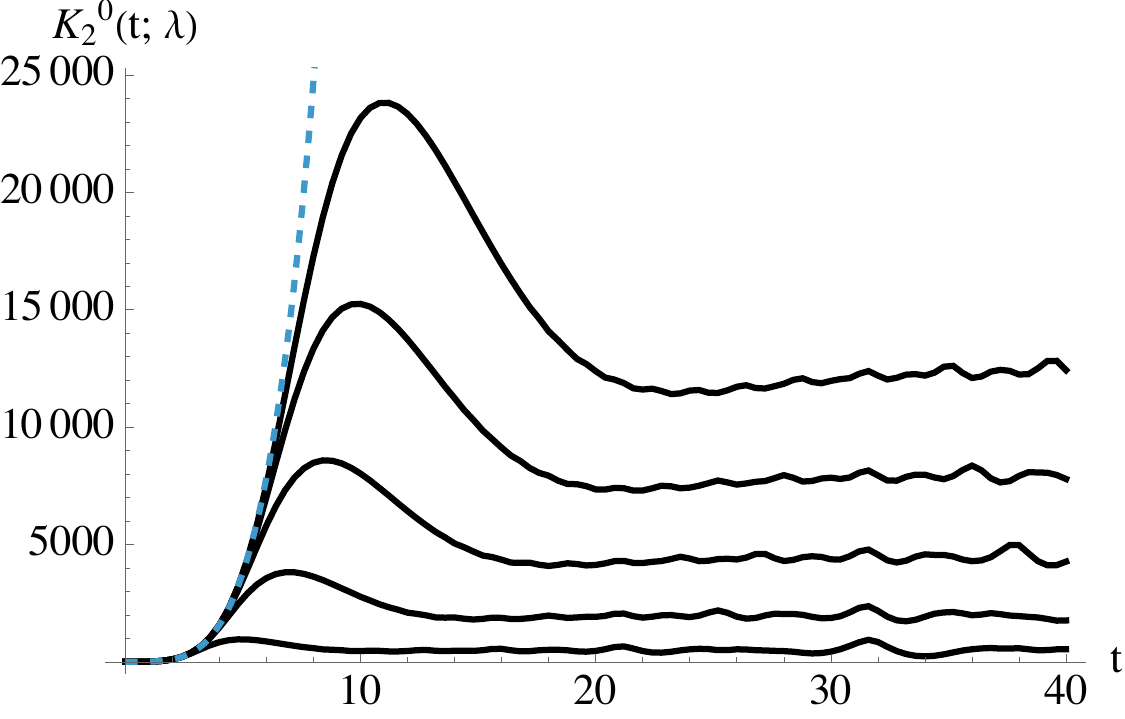}
        \caption{$\l=1$}
    \end{subfigure}

    \caption{
    Analysis of the sector $F=0$ for $M=2$. The curves in panel (a) for $\l=2/3$ correspond to $\K=50,100,150,200$, while the dashed line represents the exact result (\ref{6.6}).
    The curves in panel (b) for $\l=1$ correspond to $\K=50,100,150,200,250$, and the enveloping dashed line is the second expression in 
    (\ref{6.12}), \ie the quartic function $2t^{2}(1+3t^{2})$.
    }
    \label{fig:Mod-K02}
\end{figure}

Let us begin with the sector $F=0$, shown in Fig.~\ref{fig:Mod-K01} for $M=1$ and Fig.~\ref{fig:Mod-K02} for $M=2$. In both cases, the left panel illustrates convergence at $\l=2/3$ (a representative value smaller than $1$) as $\K$ increases. The dashed line corresponds to the exact result in (\ref{6.6}). The right panel shows the result at $\l=1$ for increasing $\K$. The dashed line represents the exact polynomial expression given in (\ref{6.12}).

\begin{figure}[htbp]
    \centering
    \begin{subfigure}{0.29\linewidth}
        \centering
        \includegraphics[width=\linewidth]{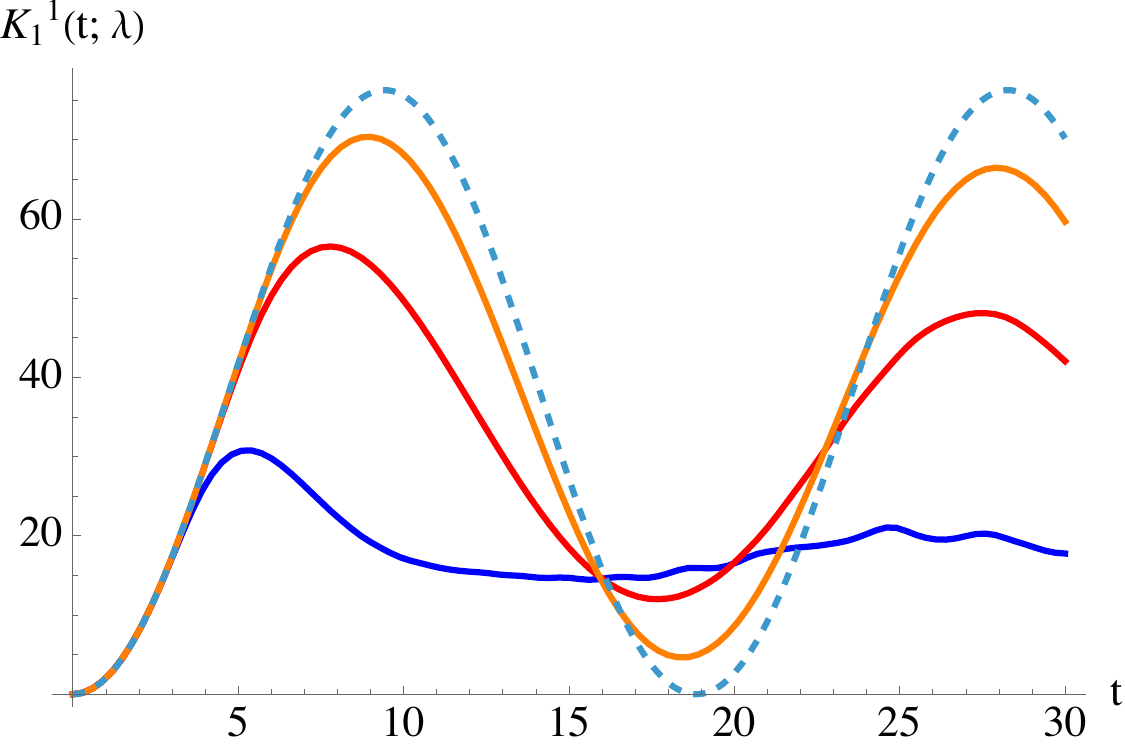}
        \caption{$\l=2/3$}
    \end{subfigure}
    \qquad
    \begin{subfigure}{0.29\linewidth}
        \centering
        \includegraphics[width=\linewidth]{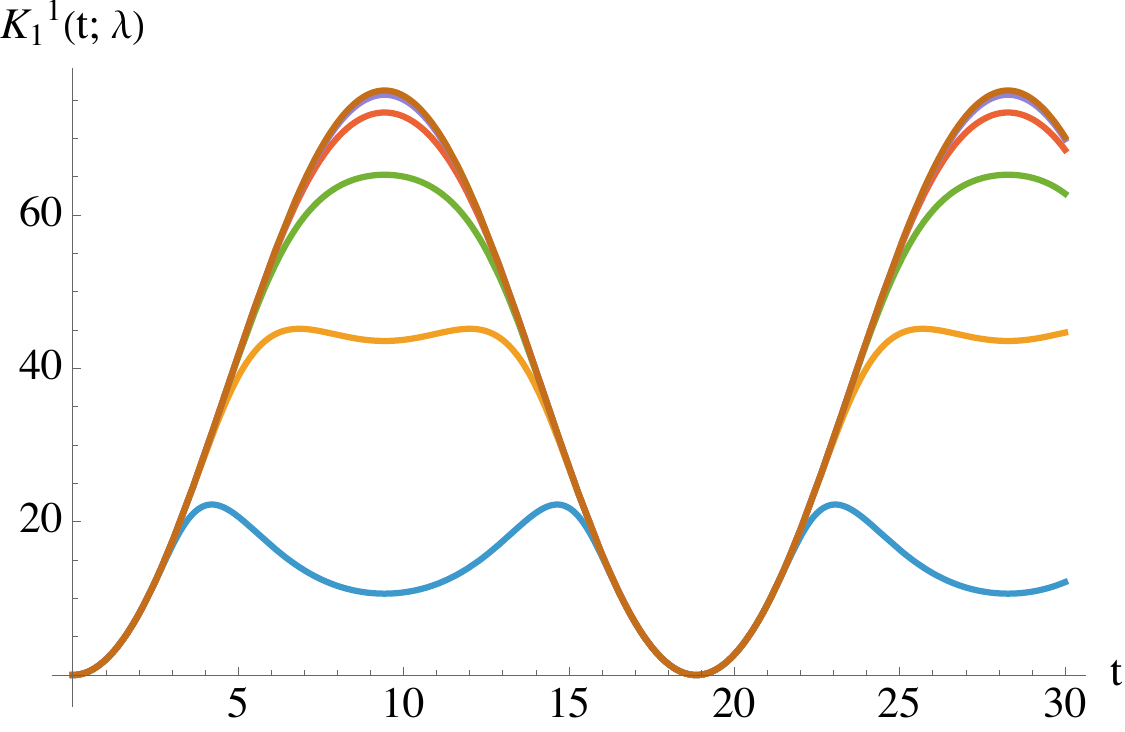}
        \caption{$\l=2/3$}
    \end{subfigure}
 \qquad
    \begin{subfigure}{0.29\linewidth}
        \centering
        \includegraphics[width=\linewidth]{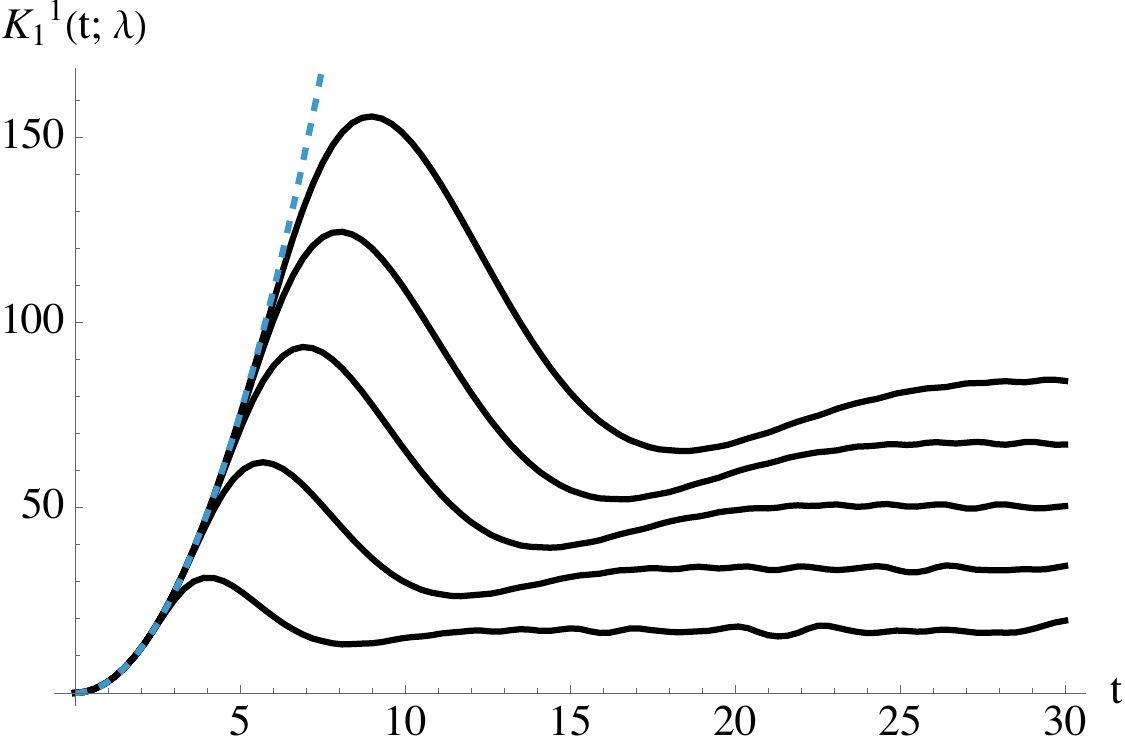}
        \caption{$\l=1$}
    \end{subfigure}

    \caption{
    Analysis of the $F=1$ sector for $M=1$. The curves in panel (a) for $\l=2/3$ correspond to $\K=50,100,150$, with the dashed line representing the sum in (\ref{6.25}) truncated to $\sim 300$ terms. 
    Convergence with the number of terms is illustrated in the middle panel (b), where the number of quasi-harmonics increases from 50 to 300 in steps of 50.
    The curves in the right panel (c) for $\l=1$ correspond to $\K=50,100,150,200,250$, while the dashed line is the exact result given in the first line of (\ref{6.43}), \ie the quadratic function $3t^{2}$.
    }
    \label{fig:Mod-K11}
\end{figure}

\begin{figure}[htbp]
    \centering
    \begin{subfigure}{0.29\linewidth}
        \centering
        \includegraphics[width=\linewidth]{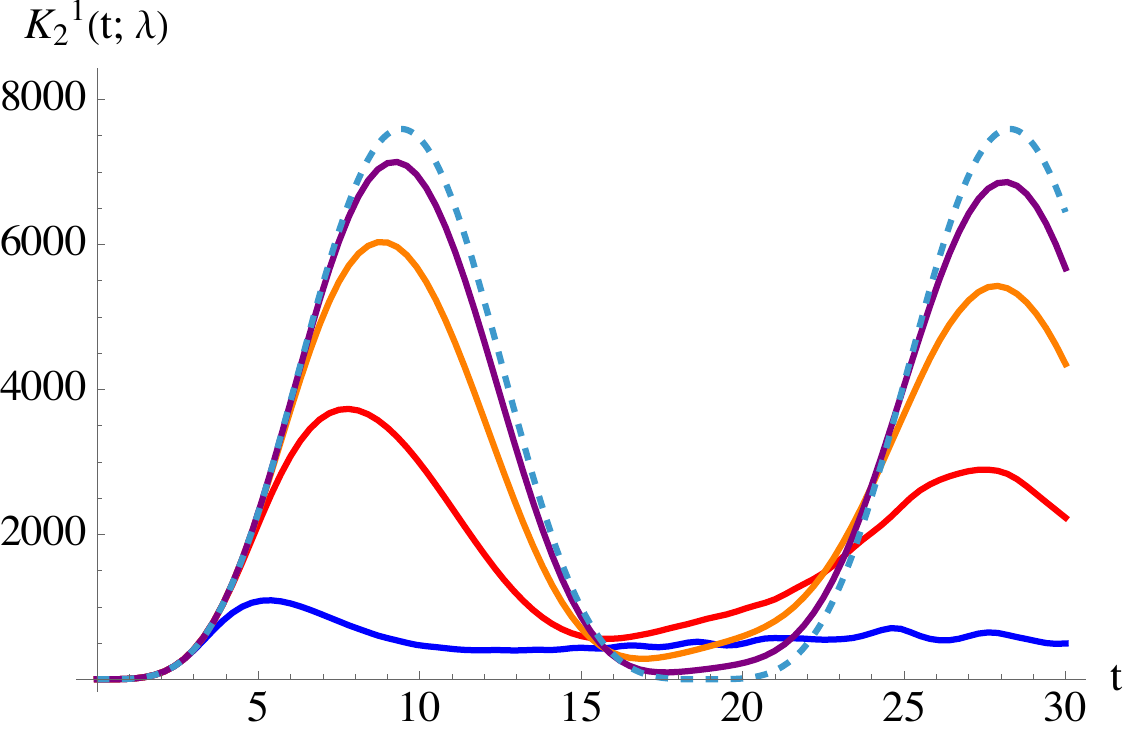}
        \caption{$\l=2/3$}
    \end{subfigure}
    \qquad
    \begin{subfigure}{0.29\linewidth}
        \centering
        \includegraphics[width=\linewidth]{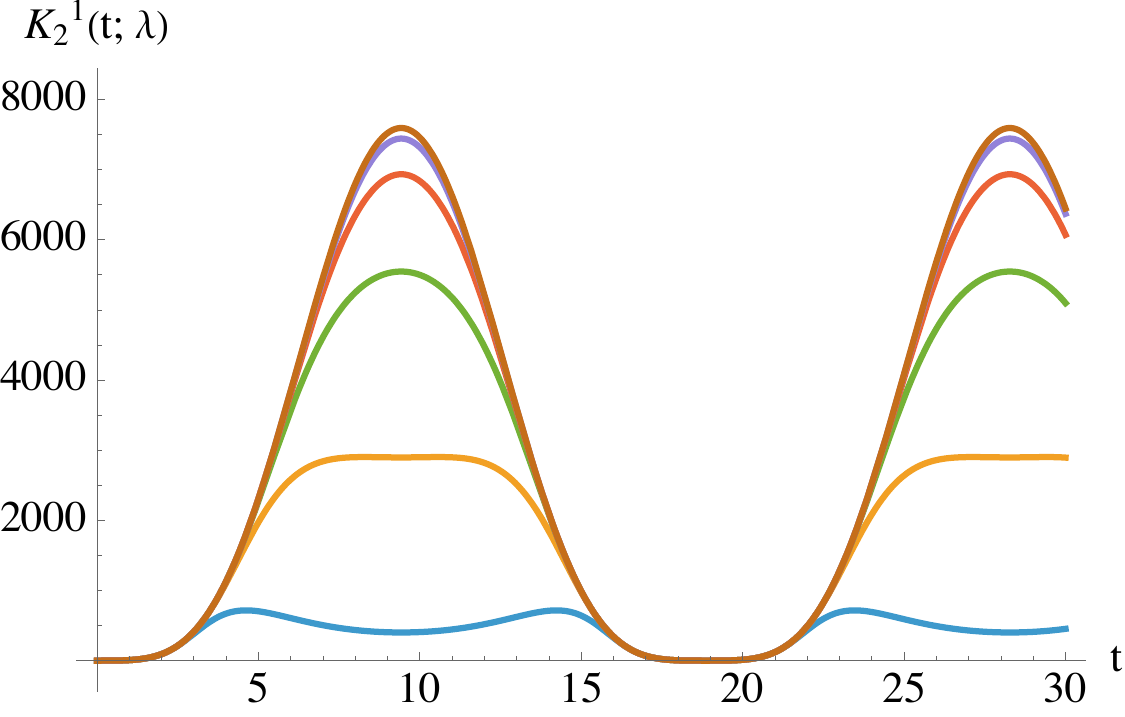}
        \caption{$\l=2/3$}
    \end{subfigure}
 \qquad
    \begin{subfigure}{0.29\linewidth}
        \centering
        \includegraphics[width=\linewidth]{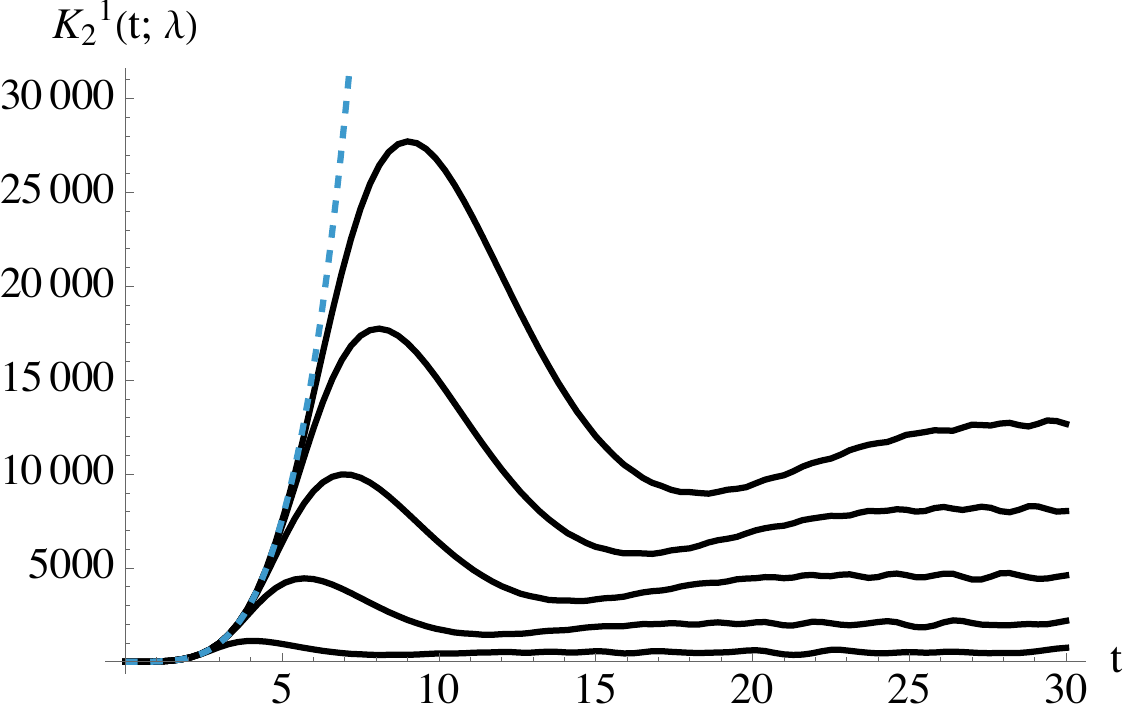}
        \caption{$\l=1$ }
    \end{subfigure}

    \caption{
    Analysis of the $F=1$ sector for $M=2$. The curves in panel (a) for $\l=2/3$ correspond to $\K=50,100,150,200$, with the dashed line representing the sum in (\ref{6.25}) truncated to $\sim 300$ terms. 
    Convergence with the number of terms is illustrated in the middle panel (b), where the number of quasi-harmonics increases from 50 to 300 in steps of 50.
    The curves in the right panel (c) for $\l=1$ correspond to $\K=50,100,150,200,250$, while the dashed line is the exact result given in the second line of (\ref{6.43}), \ie the quartic function $3t^{2}(1+4t^{2})$.
    }
    \label{fig:Mod-K12}
\end{figure}

The sector $F=1$ is shown in Figures \ref{fig:Mod-K11} for $M=1$ and \ref{fig:Mod-K12} for $M=2$. The left panel (a) again illustrates convergence at $\l=2/3$ as $\K$ increases. The dashed line corresponds to the sum in (\ref{6.25}) truncated to a large number ($\sim 300$) of terms. Convergence with respect to the number of ``harmonics'' is shown in the middle panel (b). The right panel (c) displays the result at $\l=1$, again for increasing $\K$. The dashed line represents the exact polynomial expression given in (\ref{6.43}).

\subsection{Exact inequalities}

We can prove the following exact inequalities,  holding at all times and for any $M$, relating the Krylov complexities at critical 
coupling in the bosonic and fermionic sectors
\be
\frac{3}{2}\le \frac{K^{F=1}_{M}(t)}{K^{F=0}_{M}(t)} \le 1+\frac{M}{2},
\ee
where the lower bound is attained for all $M$ at $t=0$, while the upper bound is saturated for all $t$ when $M=1$ and is never saturated for $M>1$.
These inequalities follow directly from the explicit representations (\ref{1.6}) together with the elementary bound
\be
\frac{3}{2}\le \frac{k+2}{2}\le 1+\frac{M}{2}, \qquad k=1, \dots, M.
\ee
The representations (\ref{1.6}) also show that both complexities are monotone increasing functions of time for $t>0$.
They also satisfy the relation
\be
K_{M}^{F=1}(t; 1) = \frac{1}{4t^{3}}\frac{d}{dt}[t^{4}K_{M}^{F=0}(t; 1)].
\ee
From (\ref{6.19}) and (\ref{6.41}) we obtain the large $t$ expansion of the ratio
\ba
\frac{K^{F=1}_{M}(t)}{K^{F=0}_{M}(t)} &= \frac{3t^{2}[\frac{1}{6}(M+2)!\, t^{2(M-1)}+\frac{1}{12}(M+1)! M(M-1)\, t^{2(M-2)}+\cdots]}
{2t^{2}[\frac{1}{2}(M+1)!\, t^{2(M-1)}+\frac{1}{4}M!\, M(M-1)\, t^{2(M-2)}+\cdots]}\lp
= 1+\frac{M}{2}-\frac{M(M-1)}{4(M+1)}\frac{1}{t^{2}}+\cdots,
\ea
as well as its small $t$ expansion
\be
\frac{K^{F=1}_{M}(t)}{K^{F=0}_{M}(t)} = \frac{3}{2}[1+(2^{M-1}-1)\, t^{2}+\cdots]. 
\ee
Thus the ratio interpolates between the lower bound $\frac{3}{2}$ at early times and the upper bound 
$1+\frac{M}{2}$ at late times.

\section{Complexity at $\l=1$ from the continuous spectrum}
\la{sec:cont}

In Section \ref{sec:mod-F0} we obtained the Krylov complexity in the $F=0$ sector for generic coupling $\lambda$ and then derived the $\lambda=1$ limit, see (\ref{6.12}).
In this section we show how the same result can be obtained by working directly at $\lambda=1$. 
At this critical point the discrete spectrum collapses, and the corresponding eigenstates are expected to belong to the continuous spectrum.
Their form can be read off from (\ref{5.19}), which gives
\be
\psi_{n}(E) = (-1)^{n}\frac{1}{\sqrt n}L_{n-1}^{(1)}(E).
\ee
From basic properties of associated Laguerre polynomials we have the orthogonality and completeness relations  
\be
\int_{0}^{\infty}dE E e^{-E}\ \psi_{n}(E)\psi_{m}(E) = \delta_{nm}, \qquad \sum_{n=1}^{\infty}\psi_{n}(E)\psi_{n}(E') = E^{-1}e^{E}\delta(E-E').
\ee
It is convenient to redefine the normalized eigenvectors as
\be
\braket{n}{E} = \Psi_{n}(E) \equiv \sqrt{E}e^{-E/2}(-1)^{n}\frac{1}{\sqrt n}L_{n-1}^{(1)}(E),
\ee
with
\be
\int_{0}^{\infty}dE \ \Psi_{n}(E)\Psi_{m}(E) = \delta_{nm}, \qquad \sum_{n=1}^{\infty}\Psi_{n}(E)\Psi_{n}(E') = \delta(E-E').
\ee
To compute the Krylov complexity we need a modified completeness relation with an insertion of the factor $n$, corresponding to the Krylov index.
This modified sum can be obtained from the differential equation satisfied by the associated Laguerre polynomials,
\be
\mc D_{0} L_{n-1}^{(1)}(E)= n L_{n-1}^{(1)}(E), \qquad \mc D_{0} = -E\frac{d^{2}}{dE^{2}}+(E-2)\frac{d}{dE}+1 .
\ee
It follows that for the normalized eigenstates one has 
\be
\mc D \Psi_{n}(E) = n \Psi_{n}(E), \qquad \mc D = -E\frac{d^{2}}{dE^{2}}-\frac{d}{dE}+\frac{1+E^{2}}{4E},
\ee
and therefore the matrix element of the Krylov index operator is
\be
\mmm{E}{\wh n}{E'} = \sum_{n=1}^{\infty}n \Psi_{n}(E)\Psi_{n}(E') = \mc D \delta(E-E'). %\qquad \mc D = -E\frac{d^{2}}{dE^{2}}+(E-2)\frac{d}{dE}+1 .
\ee
Using this relation it is straightforward to compute the Krylov complexity at $\lambda=1$.
One has 
\ba
K^{F=0}(t) &= 
\mmm{1}{e^{itH}(\wh n-1) e^{-itH}}{1} = \int_{0}^{\infty}dE\, dE'\, \Psi_{1}(E)\Psi_{1}(E')\, e^{it(E-E')}\, \mmm{E}{\wh n-1}{E'}\lp
= \int_{0}^{\infty} dE dE' \ \sqrt{EE'}e^{-1/2(E+E')+it(E-E')}\mc D\delta(E-E')-1.
\ea
Integrating by parts the differential operator $\mathcal D$ and then using the delta function to set $E'=E$, we obtain
\be
K^{F=0}(t)  = \int_{0}^{\infty} dE\, E\, e^{-E}\, (1+it(E-2)+E t^{2}) -1= 2t^{2},
\ee
in agreement with the first line in (\ref{6.12}). The cases with $M>1$ can be treated in the same way.

\section{Complexity of localized states in the harmonic oscillator}
\la{sec:HO}

The oscillating behaviour of the Krylov complexity at $\l\neq 1$ and the remarkably simple structure in (\ref{1.6})
at the critical point $\l=1$ are closely related to what one finds in a very elementary model, namely the one-dimensional
quantum harmonic oscillator with a Gaussian initial state. As we illustrate, the $\l\to 1$ limit corresponds (in a sense to be made precise)
to the free limit of vanishing angular frequency $\omega\to 0$ in the harmonic oscillator, where the spectrum 
changes from discrete to continuous and wave packets spread ballistically. 
The case $M=1$ and $\omega=0$ was solved in \cite{Hashimoto:2023swv}. Here we consider a general $\omega$ and study the transition
as $\omega\to 0$. It is possible to exploit the  $SU(1,1)$ structure of the harmonic oscillator and the associated  
coherent states  \cite{Perelomov1986,Adhikari:2022whf,Chowdhury:2024qaj}, but for our purposes a simple direct computation will suffice.

We consider a  quantum harmonic oscillator in one dimension with unit mass and Hamiltonian $H=\frac{1}{2}p^{2}+\frac{1}{2}\omega^{2}x^{2}$ and 
choose as initial state the localized Gaussian state $\ket{K_{0}}$ with wavefunction
\be
\la{8.1}
K_{0}(x) = \braket{x}{K_{0}} = \bigg(\frac{1}{2\pi\alpha}\bigg)^{1/4}e^{-\frac{x^{2}}{4\alpha}}.
\ee
As shown in \cite{Hashimoto:2023swv}, the full Krylov chain  is independent of $\omega$ and reads
%\be
%\wt K_{n}(p) =  \braket{p}{K_{n}} = \bigg(\frac{2\alpha}{\pi}\bigg)^{1/4}\frac{1}{\sqrt{(2n)!}\,2^{n}}H_{2n}(\sqrt{2\alpha}\, p)\, e^{-\alpha p^{2}}.
%\ee
\be
K_{n}(x) = \braket{x}{K_{n}} = \frac{1}{(2\pi \alpha)^{1/4}}\frac{1}{\sqrt{(2n)!}2^{n}}H_{2n}\bigg(\frac{1}{\sqrt{2\alpha}}x\bigg)\, e^{-\frac{x^{2}}{4\alpha}}.
\ee
The state (\ref{8.1}) remains Gaussian at any time, with a time-dependent width. 
Solving the Schr\"odinger equation one finds the breathing explicit solution
\ba
K_{0}(x, t) &= \bigg(\frac{1}{2\pi\alpha}\frac{1}{(\cos\omega t+\frac{i}{2\alpha\omega}\sin\omega t)^{2}}\bigg)^{1/4}
\exp\bigg[-\frac{1}{4\alpha}\frac{\cos\omega t+2i\alpha\omega\sin\omega t}{\cos\omega t+\frac{i}{2\alpha\omega}\sin\omega t}x^{2}\bigg] \lp
= \mc N(t) \exp(-A(t)x^{2}).
\ea
This leads to the scalar product
\ba
\braket{K_{n}}{K_{0}(t)} &= 
% \frac{1}{(2\pi \alpha)^{1/4}}\frac{1}{\sqrt{(2n)!}2^{n}}\mc N(t) \int dx H_{2n}(\frac{1}{\sqrt{2\alpha}}x)\, e^{-\frac{x^{2}}{4\alpha}-A(t)x^{2}} \lp
%= \frac{1}{(2\pi \alpha)^{1/4}}\frac{1}{\sqrt{(2n)!}2^{n}}\mc N(t)\frac{1}{\sqrt{\frac{1}{4\alpha}+A}} \int dx H_{2n}(\frac{1}{\sqrt{2\alpha}}\frac{1}{\sqrt{\frac{1}{4\alpha}+A}}x)\, e^{-x^{2}}\lp
\frac{1}{(2\pi \alpha)^{1/4}}\frac{1}{\sqrt{(2n)!}2^{n}}\mc N(t)\sqrt\frac{4\alpha}{1+4\alpha A}
\int dx H_{2n}\bigg(\sqrt\frac{2}{1+4\alpha A}\, x\bigg)\, e^{-x^{2}}.
\ea
Using
\be
\int_{-\infty}^{\infty}dx H_{2n}(ax)e^{-x^{2}} = \sqrt\pi\frac{(2n)!}{n!}(a^{2}-1)^{n},
\ee
we obtain 
\ba
\braket{K_{n}}{K_{0}(t)} &=  \frac{1}{(2\pi \alpha)^{1/4}}\frac{1}{\sqrt{(2n)!}2^{n}}\mc N(t)\sqrt\frac{4\alpha}{1+4\alpha A}
\sqrt\pi\frac{(2n)!}{n!}\bigg(\frac{1-4\alpha A}{1+4\alpha A}\bigg)^{n}.
\ea
After some simplification, the squared modulus is 
\ba
\mc P_{n}\equiv |\braket{K_{n}}{K_{0}(t)}|^{2} &=  
%\frac{1}{(2\pi \alpha)^{1/2}}\frac{1}{(2n)! 2^{2n}}|N(t)|^{2}\sqrt{\frac{4\alpha}{1+4\alpha A}\frac{4\alpha}{1+4\alpha A^{*}}}
%\pi\frac{(2n)!^{2}}{n!^{2}}\bigg(\frac{1-4\alpha A}{1+4\alpha A}\frac{1-4\alpha A^{*}}{1+4\alpha A^{*}}\bigg)^{n} \lp
%= \frac{\pi}{(2\pi \alpha)^{1/2}}\frac{(2n)!}{n!^{2} 2^{2n}}
%\sqrt\frac{2\alpha}{\pi}\omega\frac{1}{\sqrt{4\alpha^{2}\omega^{2}\cos^{2}\omega t+\sin^{2}\omega t}}\lp
%4\alpha\frac{\sqrt{4\alpha^{2}\omega^{2}\cos^{2}\omega t+\sin^{2}\omega t}}{\sqrt{16\alpha^{2}\omega^{2}\cos^{2}\omega t+(1+4\alpha^{2}\omega^{2})^{2}\sin^{2}\omega t}}
%\bigg(\frac{(1-4\alpha^{2}\omega^{2})^{2}\sin^{2}\omega t}{16\alpha^{2}\omega^{2}\cos^{2}\omega^{2}+(1+4\alpha^{2}\omega^{2})^{2}\sin^{2}\omega t}\bigg)^{n}\lp
%= \frac{4\alpha\omega}{\sqrt{16\alpha^{2}\omega^{2}\cos^{2}\omega t+(1+4\alpha^{2}\omega^{2})^{2}\sin^{2}\omega t}}\frac{(2n)!}{n!^{2} 2^{2n}}
%\bigg(\frac{(1-4\alpha^{2}\omega^{2})^{2}\sin^{2}\omega t}{16\alpha^{2}\omega^{2}\cos^{2}\omega^{2}+(1+4\alpha^{2}\omega^{2})^{2}\sin^{2}\omega t}\bigg)^{n}\lp
%= \frac{4\alpha \omega}{\sqrt{16\alpha^{2}\omega^{2}+(1-4\alpha^{2}\omega^{2})^{2}\sin^{2}\omega t}}\frac{(2n)!}{n!^{2} 2^{2n}}
%\bigg(\frac{(1-4\alpha^{2}\omega^{2})^{2}\sin^{2}\omega t}{16\alpha^{2}\omega^{2}+(1-4\alpha^{2}\omega^{2})^{2}\sin^{2}\omega t}\bigg)^{n}\lp
%= \frac{1}{\sqrt{1+\frac{(1-4\alpha^{2}\omega^{2})^{2}}{16\alpha^{2}\omega^{2}}\sin^{2}\omega t}}\frac{(2n)!}{n!^{2} 2^{2n}}
%\bigg(1- \frac{1}{1+\frac{(1-4\alpha^{2}\omega^{2})^{2}}{16\alpha^{2}\omega^{2}}\sin^{2}\omega t}\bigg)^{n}\lp
\sqrt{1-Y(t)}\frac{(2n-1)!!}{n! 2^{n}}Y(t)^{n},\qquad Y(t) = 1-\frac{1}{1+\frac{(1-4\alpha^{2}\omega^{2})^{2}}{16\alpha^{2}\omega^{2}}\sin^{2}\omega t}.
\ea
Following \cite{Hashimoto:2023swv} we note that
\be
\frac{1}{\sqrt{1-y}} = \sum_{n=0}^{\infty}\frac{(2n-1)!!}{n!2^{n}}y^{n}.
\ee
This implies that ($Y\equiv Y(t)$)
\ba
\la{8.11}
K_{M}^{\rm HO}(t; \omega) = \sum_{n=1}^{\infty}n^{M}\,\mc P_{n} = \sqrt{1-Y}\left(Y\frac{\partial}{\partial Y}\right)^{M}\frac{1}{\sqrt{1-Y}}.
\ea
Let us introduce the variable $T$ defined by 
\be
Y = \frac{T^{2}}{1+T^{2}}, \qquad T = \frac{|1-4\alpha^{2}\omega^{2}|}{4\alpha\omega}\sin\omega t.
\ee
Then  (\ref{8.11}) can be written as 
\ba
\la{8.11}
K_{M}^{\rm HO}(t; \omega) = \frac{1}{\sqrt{1+T^{2}}}\left(\frac{1}{2}T(1+T^{2})\frac{\partial}{\partial T}\right)^{M}\sqrt{1+T^{2}}.
\ea
Using the results in Appendix \ref{app:poly},  in particular  (\ref{C.7}) and (\ref{C.11}),  one obtains the explicit expression
\ba
\la{8.12}
K_{M}^{\rm HO}(t; \omega) &=\frac{1}{\sqrt{\pi}}\sum_{k=1}^{M} S_{2}(M,k)\Gamma\left(k+\frac{1}{2}\right)\, \bigg(\frac{(1-4\alpha^{2}\omega^{2})^{2}}{16\alpha^{2}\omega^{2}}\sin^{2}\omega t\bigg)^{k}.
\ea
The first cases are 
\ba
K_{1}^{\rm HO}(t; \omega) &=\frac{(1-4\alpha^{2}\omega^{2})^{2}}{32\alpha^{2}\omega^{2}}\sin^{2}\omega t,\\
K_{2}^{\rm HO}(t; \omega) &=\frac{(1-4\alpha^{2}\omega^{2})^{2}}{32\alpha^{2}\omega^{2}}\sin^{2}\omega t
+\frac{3(1-4\alpha^{2}\omega^{2})^{4}}{1024\alpha^{4}\omega^{4}}\sin^{4}\omega t,
\ea
and so on. The complexity $K_{M}^{\rm HO}(t; \omega)$ oscillates in time, reflecting the equally spaced spectrum of the harmonic oscillator. \footnote{
If we take $\alpha=1/(2\omega)$ in (\ref{8.12}), i.e. we start with the ground state, the complexity vanishes because $K_{0}(t)$ coincides with  $K_{0}$ up to a phase.
}
Taking the free particle limit $\omega\to 0$ one obtains
\ba
K_{M}^{\rm HO}(t; 0) &=\frac{1}{\sqrt{\pi}}\sum_{k=1}^{M}S_{2}(M,k)\Gamma\left(k+\frac{1}{2}\right)\left(\frac{t}{4\alpha}\right)^{2k},
\ea
with the first cases
\ba
K_{1}(t) &= \frac{t^{2}}{2^{5}\alpha^{2}}, \qquad 
K_{2}(t) = \frac{t^{2}}{2^{10}\alpha^{4}}(3\,t^{2}+32\alpha^{2}), \qquad
K_{3}(t) = \frac{t^{2}}{2^{15}\alpha^{6}}(15\,t^{4}+288\alpha^{2}\,t^{2}+1024\alpha^{4}), \nonumber \\
K_{4}(t) &= \frac{t^{2}}{2^{20}\alpha^{8}}(105\,t^{6}+2880\alpha^{2}\,t^{4}+21504\alpha^{4}\,t^{2}+32768\alpha^{6}).
\ea
In conclusion, for $\omega\neq 0$ the complexity oscillates in time, while in the free limit $\omega=0$ it grows polynomially as $\sim t^{2M}$ 
with subleading terms governed by Stirling numbers of the second kind. This behaviour closely parallels what we observe in the Veneziano--Wosiek model 
in the two regimes $\lambda\neq 1$ and $\lambda=1$.

\section*{Acknowledgements}
We thank Hyun-Sik Jeong for useful comments.
MB is supported by the INFN grant GAST. EA is supported by  
the  MUR project GINEVRA, prot. 2022BZYBWM.

\appendix

%
%\section{Recovering the Veneziano--Wosiek $F=0$ spectrum from the modified Hamiltonian}
%\la{app:rank1}
%
%As discussed in \cite{DePietri:2006hi}, 
%the relation between $H$ in (\ref{5.1}) and $\wt H$ in (\ref{5.2}) is a rank-1 perturbation of the form
%\be
%H = \wt H-\l \ket{1}\bra{1}.
%\ee
%This implies that one can recover the spectrum of $H$ from that of $\wt H$ by using the identity
% \be
% \la{A.2}
% \det(M+uv^{T}) = \det M\, (1+v^{T}M^{-1}u).
% \ee
% Let us focus for the purpose of illustration the case $0<\l<1$. From (\ref{A.2}), we see that the eigenvalues of $H$ are given by the equation
% \be
% 1-\l\mmm{1}{(\wt H-E)^{-1}}{1}=0 \qquad\to \qquad
% 1-\l\sum_{p=1}^{\infty}\frac{\braket{1}{\psi^{(p)}}^{2}}{(1-\l)p-E}=0.
% \ee
% From (\ref{5.10})
% \be
% \braket{1}{\psi^{(p)}}^{2} = p(1-\l)^{2}\l^{1-p}F(-(p-1), 2, 2, 1-\l)^{2} = p(1-\l)^{2}\l^{p-1}.
% \ee
% Computing the sum and simplifying this gives the quantization condition in  (\ref{3.26})
% \be
% B\bigg(\l, \frac{E}{\l-1},0\bigg)=0,
% \ee
% where we recall that $B(z, a,b)$ denotes the incomplete beta function. 
 
 \section{Computation of $|\mc B_{n}(t; \l)|^{2}$}
 \la{app:sum}

Taking into account (\ref{5.10}), the sum (\ref{6.2}) reads ($\mu \equiv (1-\l)\, t$)
\ba
\la{B.1}
\mc B_{n}(t; \l) &= \sum_{p=1}^{\infty}e^{-i\mu p}\psi^{(p)}_{1}\psi^{(p)}_{n} 
= \sum_{p=1}^{\infty}(-1)^{n+1}\sqrt{n}\l^{\frac{n-1}{2}}(1-\l)^{2}e^{-i\mu p}\, \FF(n+1, 1-p, 2; 1-\l)\lp
= (-1)^{n+1}\sqrt{n}\l^{\frac{n-1}{2}}(1-\l)^{2} G_{n+1}(e^{-i\,\mu}, 1-\l),
\ea
where we introduced
\be
G_{n}(x,y) = \sum_{p=1}^{\infty}p\, x^{p}\,\FF(n,1-p,2,y).
\ee
Since $n,p$ are positive integers we have 
\ba
\la{B.3}
G_{n}(x,y) &= \sum_{p=1}^{\infty}p\, x^{p}\, \sum_{k=0}^{p-1}(-1)^{k}\frac{(n)_{k}}{k!(k+1)!}\frac{(p-1)!}{(p-1-k)!}y^{k}\lp
= \sum_{k=0}^{\infty}(-1)^{k}\frac{(n)_{k}}{k!(k+1)!}y^{k}\, \sum_{p=k+1}^{\infty}p\, x^{p}\,\frac{(p-1)!}{(p-1-k)!},
\ea
where we exchanged summations. The inner sum is 
\ba
\sum_{p=k+1}^{\infty}p\, x^{p}\,\frac{(p-1)!}{(p-1-k)!} &= \sum_{p=0}^{\infty}(p+k+1)\, x^{p+k+1}\,\frac{(p+k)!}{p!} = x^{k+1}\sum_{p=0}^{\infty}\frac{(p+k+1)!}{p!}x^{p}\lp
= (k+1)!\frac{x^{k+1}}{(1-x)^{k+2}}.
\ea
Inserting this into (\ref{B.3}), we get 
\ba
G_{n}(x,y) &= 
\sum_{k=0}^{\infty}(-1)^{k}\frac{(n)_{k}}{k!(k+1)!}y^{k}\, (k+1)!\frac{x^{k+1}}{(1-x)^{k+2}}
= \sum_{k=0}^{\infty}(-1)^{k}\binom{k+n-1}{k}y^{k}x^{k+1}(1-x)^{-2-k} \lp
= \frac{x}{(1-x)^{2}}\bigg(1+\frac{xy}{1-x}\bigg)^{-n}.
\ea
Using this in (\ref{B.1}) we obtain 
\be
\mc B_{n}(t; \l) = (-1)^{n+1}\sqrt n\, (1-\l)^{2}\l^{\frac{n-1}{2}}\frac{e^{i\mu}}{(e^{i\mu}-1)^{2}}\, \bigg(\frac{e^{i\mu}-1}{e^{i\mu}-\l}\bigg)^{n+1}.
\ee
Taking the squared modulus, after some simplification, we obtain  $|\mc B_{n}(t; \l)|^{2}$ as in  (\ref{6.3}).

\section{A class of polynomials relevant at the $\l=1$ critical point}
\la{app:poly}

Let us introduce the polynomials
\be
\mc P_{M}(t; \alpha) = \bigg[1+\alpha t^{2}+\frac{1}{2}t(1+t^{2})\frac{d}{dt}\bigg]\,\mc P_{M-1}(t; \alpha), \qquad \mc P_{1}(t; \alpha) = 1,
\ee
where $\alpha$ is a constant parameter. The first cases are 
\bea
\mc P_{2}(t; \alpha) &= 1+\alpha\, t^{2}, \\
\mc P_{3}(t; \alpha) &= 1+3\alpha\, t^{2}+\alpha(1+\alpha)\, t^{4}, \\
\mc P_{4}(t; \alpha) &= 1+7\alpha\, t^{2}+6\alpha(1+\alpha)\, t^{4}+\alpha(1+\alpha)(2+\alpha)\, t^{6}.
\eea
Changing variables to 
\be
x = \frac{t^{2}}{1+t^{2}},
\ee
and denoting $\PP_{M}(x(t); \alpha) =  \mc P_{M}(t; \alpha)$, we have 
\be
\PP_{M}(x; \alpha) = \bigg[1+\alpha\frac{x}{1-x}+x\frac{d}{dx}\bigg]\,\PP_{M-1}(x; \alpha), \qquad \PP_{1}(x; \alpha) = 1,
\ee
Let us define
\be
\PP_{M}(x; \alpha) = \frac{(1-x)^{\alpha}}{x}\, \bar{\mc P}_{M}(x; \alpha).
\ee
We get 
\be
\bar{\mc P}_{M}(x; \alpha)= x\frac{d}{dx}\bar{\mc P}_{M-1}(x; \alpha), \qquad \bar{\mc P}_{1}(x; \alpha) = \frac{x}{(1-x)^{\alpha}}.
\ee
Thus, 
\ba
\la{C.7}
\wt{\mc P}_{M}(x; \alpha) &=\frac{(1-x)^{\alpha}}{x}\bigg(x\frac{d}{dx}\bigg)^{M-1} \frac{x}{(1-x)^{\alpha}}.
\ea
The Stirling number of second kind $S_{2}(M,N)$
\be
\la{C.8}
S_{2}(M,N) = \frac{1}{N!}\sum_{n=0}^{N}(-1)^{N-n}\binom{N}{n}n^{M},
\ee
obey the relation
\be
\sum_{n=k}^{\infty}S(n,k)\frac{x^{n}}{n!} = \frac{1}{k!}(e^{x}-1)^{k}.
\ee
Changing variable in (\ref{C.7}) gives then the final explicit formula valid for $M=2, 3, \dots$
\be
\mc P_{M}(t; \alpha) = \sum_{k=1}^{M-1}S_{2}(M-1,k)\frac{\Gamma(k+\alpha-1)}{\Gamma(\alpha)}[k+(k+\alpha-1)\,t^{2}]t^{2(k-1)},
\ee
that we may also write as (notice that $k$ runs up to $M$ now) \footnote{We use the recursion $S_{2}(m+1,k)=k S_{2}(m,k)+S_{2}(m,k-1)$.}
\ba
\la{C.11}
\mc P_{M}(t; \alpha) &=\sum_{k=1}^{M}S_{2}(M,k)\,\frac{\Gamma(k+\alpha-1)}{\Gamma(\alpha)}t^{2(k-1)}
\ea
Using $S_{2}(M-1,M)=0$, the leading term of maximal degree is 
\be
\la{C.12}
\mc P_{M}(t; \alpha) = \frac{\Gamma(M+\alpha-1)}{\Gamma(\alpha)}\, t^{2(M-1)}+\mc O(t^{2(M-2)}).
\ee

\section{Krylov complexity with a generic initial state}
\la{app:generic}

\begin{figure}[htbp]
    \centering

    \begin{subfigure}{0.4\linewidth}
        \centering
        \includegraphics[width=\linewidth]{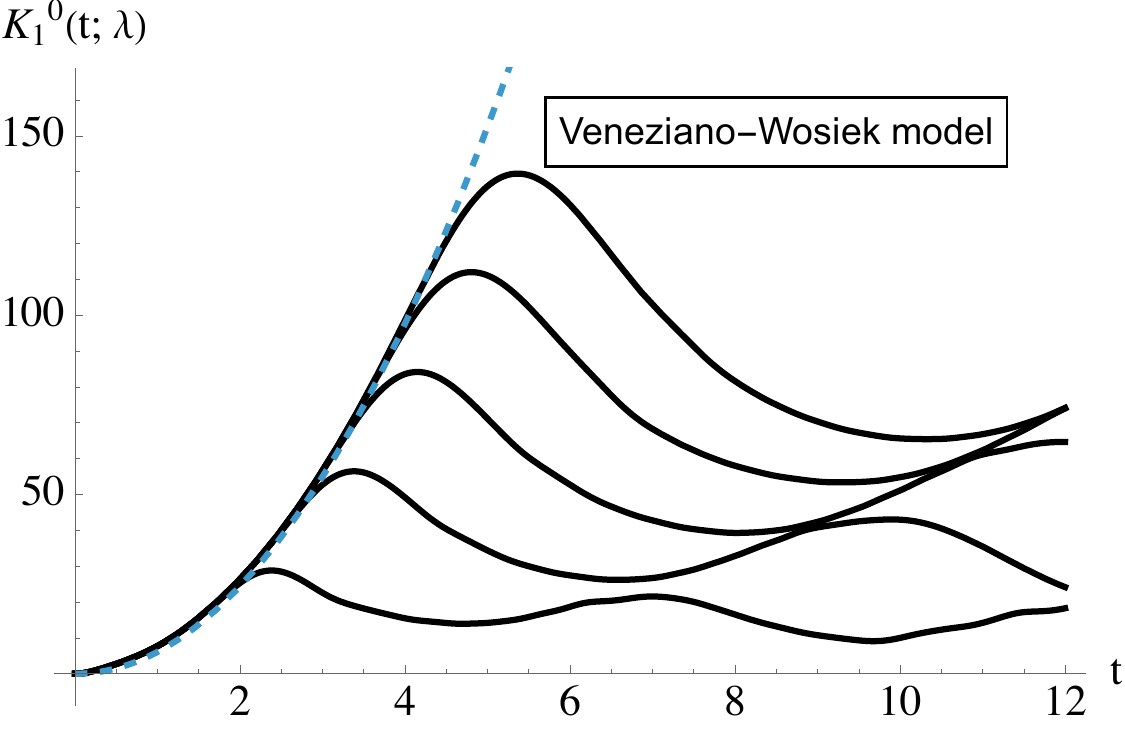}
        \caption{}
%        \label{fig:sub1}
    \end{subfigure}
    \qquad
    \begin{subfigure}{0.4\linewidth}
        \centering
        \includegraphics[width=\linewidth]{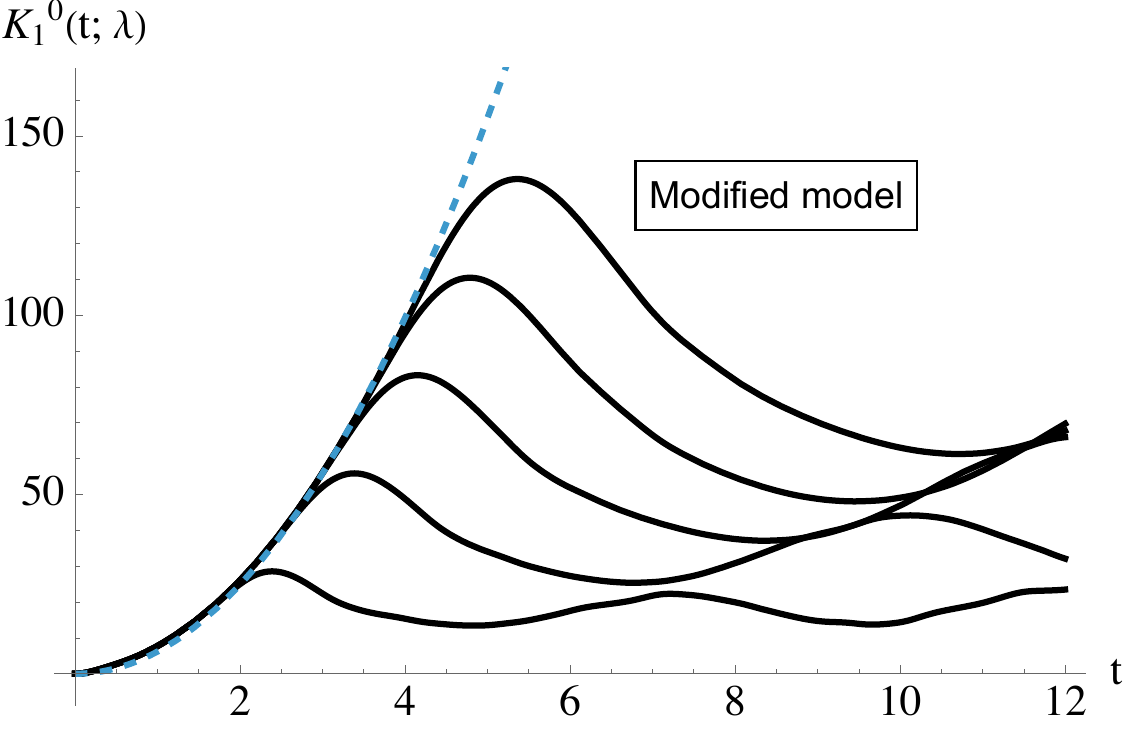}
        \caption{}
%        \label{fig:sub2}
    \end{subfigure}

    \caption{
    Analysis of the sector $F=0$ for $M=1$ and initial state $\ket{0;3}$. 
    In both panels the curves correspond to $\lambda=1$ and $\K=50,100,150,200,250$ (from bottom to top).
    The dashed lines are guides to the eye and correspond to the quadratic function $6.2\,t^{2}$, which provides a good approximation in both models
    showing that for this initial state, the rank-1 modification produces very small deviations. 
    }
    \label{fig:excited}
\end{figure}

So far we have considered the Krylov complexity starting from a special state, chosen so as to exploit the 
tridiagonal form of the Hamiltonian in the orthonormal basis of large-$N$ occupation-number states.
To briefly illustrate what happens for a generic initial state, we consider the case $s_{0}=\ket{0;3}$ in the $F=0$ sector, both in the 
Veneziano--Wosiek and in the modified models.
The corresponding numerical results are shown in Figure \ref{fig:excited}.
At $\lambda=1$ the complexity is quadratic in time in both models, with a normalization that depends on the initial state.
This simple example suggests that the quadratic growth at $\lambda=1$ is robust with respect to the 
choice of the initial state, although the overall normalization of the complexity depends on the state. For a recent discussion of the
dependence of Krylov complexity on the initial state, see \cite{PG:2025ixk}.

 \bibliography{BT-Biblio}
\bibliographystyle{JHEP-v2.9}
\end{document}